\documentclass{aa}

\DeclareRobustCommand{\disambiguate}[3]{#2~#3}
\usepackage{amsmath, amsthm, amssymb, amsfonts}
\usepackage{txfonts}
\usepackage{graphicx}
\usepackage{natbib}
\usepackage{subfig}
\usepackage{siunitx}
\usepackage{cleveref}
\usepackage{placeins}
\usepackage{xcolor}
\crefname{section}{§}{§§}
\Crefname{section}{§}{§§}
\crefformat{section}{§#2#1#3}
\bibpunct{(}{)}{;}{a}{}{,} 

\begin{document}
        
           \title{Warp and flare of the Galactic disc revealed with supergiants by Gaia EDR3}
        
        \subtitle{}
        
        \author{\v{Z}. Chrob\'{a}kov\'{a}\inst{1,2,3}, R. Nagy\inst{1,4}, M. L\'{o}pez-Corredoira\inst{2,3}}
        
        \institute{Equal first authors
                        \and
                        Instituto de Astrofísica de Canarias, E-38205 La Laguna, Tenerife, Spain
                \and
                Departamento de Astrofísica, Universidad de La Laguna, E-38206 La Laguna, Tenerife, Spain
                \and
                Faculty of Mathematics, Physics, and Informatics, Comenius University, Mlynsk\'{a} dolina, 842 48 Bratislava, Slovakia
        }
        
        \date{Received xxxx; accepted xxxx}
        
        
        \abstract
        {The outer Galactic disc contains some features such as the warp and flare, whose origin is still debated. The Gaia data provide an excellent opportunity to probe the Galactic disc at large distances and study these features.}
        {We derive the density distributions of the average (old) whole population and the supergiants (representative of a young population), and we use them to constrain their warp and flare. By comparing the results, we study how the properties of these phenomena depend on the studied population.}
        {We used Lucy's deconvolution method to recover corrected star counts as a function of distance, from which we derive the density distribution.}
        {We find that supergiants have an asymmetric warp, reaching a maximum amplitude of $z_w=0.658$ kpc and minimum amplitude of $z_w=-0.717$ kpc at a distance of $R=[19.5,20]$ kpc, which is almost twice as high as the amplitude of the whole population of the disc. We find a significant flare of the whole population, especially in the thick disc. The scale height increases from $h_{z,thick}\approx 0.7$ kpc and $h_{z,thin}\approx 0.3$ kpc in the solar neighbourhood, to $h_{z,thick} \approx 2.6$ kpc and $h_{z,thin}\approx 0.6$ kpc in the remote regions of the Milky Way ($R\approx 18$ kpc). The supergiants' population has only a small flare.}
        {}
        
        \titlerunning{Warp and flare of the Galactic disc revealed with supergiants by Gaia EDR3}
        \authorrunning{\v{Z}. Chrob\'{a}kov\'{a} et al.}
        \maketitle

\section{Introduction}  

Although we have an extensive knowledge of our Galaxy, there are still many aspects for which our understanding of the Milky Way is incomplete. Wide discussions about features such as warp, flare, or cut-off show that there is much more to be learned about the Milky Way. To this end, the Gaia Early Data Release 3 \cite[EDR3]{gaia_edr3_main} presents an opportunity to study the Galaxy in greater detail than ever before. With its precise positional, proper motions, radial velocity measurements, and distance determinations for millions of stars, it offers the most accurate information about our Galaxy to date; ideal for making advances in all branches of Galactic astrophysics.

The Galactic warp is a well-known feature of the Galactic disc, but its shape is constrained only roughly and there is no consensus on the mechanism causing it. Some of the theories include accretion of intergalactic matter onto the disc \citep{martin_accretion}, a misaligned halo  \citep{misalign_halo}, interaction with satellites \citep{warp_sat}, or an intergalactic magnetic field \citep{mag_fields}. Currently the kinematical information about the warp is not sufficient to constrain the formation models. \cite{zofi} compared the warp of the whole stellar population with the warp of Cepheids and suggest that warp is dependent on the age of the studied population, which would indicate that warp is caused by a non-gravitational mechanism. A similar conclusion was reached by \cite{haifeng_2020} using LAMOST DR4. In this paper, we recalculate the Galactic warp using the most recent Gaia EDR3 data and compare this with the warp of supergiants, to test this hypothesis.

The flare is an increase in the scale height of the Galactic disc with galactocentric radius, detected in both the gaseous and stellar components. \cite{grabelsky} and \cite{may} confirmed flare in the outer disc by tracing molecular clouds, \cite{sanchez} modelled the flare applying modified Newtonian dynamics (MOND), and \cite{narayan_flare} treated HI, H$_2$, and stars as gravitationally coupled components of the disc, calculating scale heights for all three components and giving predictions for the flare that matched observations very well.
Stellar flare was studied by several authors as well \citep{alard, martin_velky_clanok, momany}. \cite{martin_flare} studied 3D stellar distribution using Sloan Extension for Galactic Understanding and Exploration (SEGUE) data, finding that flare is a prominent feature for Galactocentric distance of $R \gtrsim 15$ kpc. \cite{yusifov} reached a similar conclusion studying pulsars. \cite{bovy} studied the structure of the Galactic disc, distinguishing various stellar populations using APOGEE survey data covering Galactocentric distances of $R < 15$ kpc. They did not find stellar flaring in the high-$[\alpha/Fe]$ mono-abundance population, while the low-$[\alpha/Fe]$ mono-abundance populations exhibit Galactic flare.

To explain the dependence of the disc thickness on azimuth, \cite{kalberla} modelled it with a ring of dark matter embedded in the disc, and \cite{saha} applied a lopsided dark matter halo. \cite{martin_flare_letter} proposed accretion of intergalactic matter onto the disc as a possible mechanism to explain both the flare and its dependence on azimuth.

More recently, the flare was studied with Gaia DR2 using OB stars \citep{li} or with LAMOST \citep{haifeng} and Cepheids \citep{feast}. Possible existence of the flare was also explored in the thick disc \citep{martin_flare,mateu,haifeng}. \cite{yu} studied the warp and the flare traced by OB stars using LAMOST DR5 data and \cite{yu2} investigated the Galactic disc using LAMOST and Gaia Red Clump Sample VII.

In this paper, we use the Gaia EDR3 data to study the warp and flare. We are especially interested in the properties of these features for various stellar populations, and therefore we analyse separately the population of supergiants. The paper is structured as follows. In Section 2 we describe our data selection and the extinction map used. In Section 3 we present the method used to calculate the density distribution and explain how we chose the sample of supergiants. Section 4 is dedicated to the analysis of the warp and Section 5 deals with the analysis of the flare. In Section 6 we conclude the paper.  

\section{Data Selection}\label{datasel}
We used the Gaia EDR3 data, collected during the first 34 months of observations. We are interested in sources with G-band (330–1050
nm) magnitude. The photometric uncertainties were $\sim0.3$ mmag for G<13, 1 mmag at G=17, and 6 mmag at G=20 mag. We chose a magnitude up to G=19, where the catalogue was sufficiently complete \citep{edr3_validation}. More details on the catalogue validation are available in \cite{edr3_validation}.
To ensure the quality of the dataset, we applied several constraints. To select sources with good astrometry, we only chose data with five- and six-parameter solutions, satisfying the following condition on renormalised unit weight error (RUWE): 

\begin{eqnarray}
\texttt{RUWE}<1.4 \nonumber \\
\end{eqnarray}
as suggested by \cite{lindegren_conditions}. In addition, following \cite{antoja_gaiaedr3}, we applied the following constraint to ensure that the sources had good photometry:

\begin{eqnarray}
0.01+0.039~(BP-RP)\hspace{-0.3cm}&<&\hspace{-0.3cm}log_{10}~\texttt{excess\_flux} \\
&<&0.12+0.039~ (BP-RP)~, \nonumber
\end{eqnarray} 
where $BP-RP$ denotes the colour index, with the $G_{BP}$ band covering the range 330--680 nm and the $G_{RP}$ band covering the range 640--1050 nm, and \texttt{excess\_flux} is the BP and RP flux excess, corrected as suggested by \cite{riello_kod}. Moreover, we applied:

\begin{eqnarray}
\texttt{phot\_g\_mean\_flux\_over\_error}&>&50~, \nonumber \\ 
\texttt{phot\_rp\_mean\_flux\_over\_error}&>&20~, \nonumber \\ 
\texttt{phot\_bp\_mean\_flux\_over\_error}&>&20~, \nonumber \\ 
\end{eqnarray}
which removed variable stars \citep{gaiadr2_hr}.
We followed the approach of \cite{zofi} and chose data with a parallax in the interval [0,2] mas, and apparent magnitude in the $G$- band between G=12 and G=19. We corrected $G$ fluxes for six-point sources as well, as suggested by \cite{riello_kod}, using codes listed in the appendix of \cite{gaia_edr3_main}. We also added the zero-point correction as found by \cite{lindegren_conditions}, using the publicly available Python package\footnote{\url{https://gitlab.com/icc-ub/public/gaiadr3_zeropoint}}, which calculates the zero-point as a function of ecliptic latitude, magnitude, and colour.

\subsection*{Extinction map}
In order to estimate the extinction, we used the three-dimensional, full-sky extinction map from \cite{ext_mapa_combined}, using the Python package \textit{mwdust}. This extinction map is a combination of maps of \cite{marshall}, \cite{green}, and \cite{drimmel}, and provides reddening as defined in \cite{sfd}. \\
In order to convert the interstellar reddening of these maps into $E(B-V)$, we used the coefficients \citep{hendy_ext_koef,rybizki_ext_koef}:

\begin{eqnarray}\label{1}
\begin{split}
A_G/A_v&=0.859~, \\
R_V&=A_v/E(B-V)=3.1~.
\end{split}
\end{eqnarray}

\section{Methods}\label{meth}
We followed the approach of \cite{zofi}, who used the fundamental equation of stellar
statistics \citep{chandrasekhar} to derive the stellar density:

\begin{eqnarray}\label{4}
\rho(1/\pi)&=&\frac{N(\pi)\pi^4 }{\Delta\pi\omega  \int_{M_{G,low~lim}}^{M_{G,low~lim}+1} \mathrm{d}M_G\Phi(M_G)}~,
\end{eqnarray}

\begin{eqnarray}
M_{G,low~lim}&=&m_{G,low~lim}-5log_{10}(1/\pi)-10 \nonumber \\
&-&A_{G}(1/\pi)~,
\end{eqnarray}
where $N(\pi)$ are the star counts, $\omega$ is the covered angular surface, $\Delta\pi$ is the parallax interval (0.01 mas in our case, which must be added in the equation because we did not use the unit parallax), $\Phi(M_G)$ is the luminosity function in the G filter, $m_{G,low~lim}$ is the limiting maximum apparent magnitude, and $A_{G}(r)$ is the extinction as a function of distance. For the luminosity function, we used the values given in Table 1 of \cite{zofi} for the whole population; or we determined it following the same method for some sub-samples.

The density determination required that we measure the star counts as a function of distance. However, it is well known that the parallax error grows with the distance from the observer, preventing the precise determination of distances. Therefore, we could not simply use the observed star counts to calculate the density, as at distances higher that roughly 5 kpc they are biased. In order to recover correct star counts, we applied a statistical deconvolution method developed by \cite{martin} based on Lucy's method \citep{lucy}. They express the observed number of stars per parallax $\overline{N}(\pi)$ as a convolution of the real number $N(\pi)$ of stars with a Gaussian function: 

\begin{eqnarray}\label{2}
\overline{N}(\pi)=\int_{0}^{\infty} \mathrm{d}\pi^\prime N(\pi^\prime)G_{\pi^\prime}(\pi-\pi^\prime)~,
\end{eqnarray}
where

\begin{eqnarray}\label{3}
G_{\pi}(x)=\frac{1}{\sqrt{2\pi}\sigma_\pi}e^{-\frac{x^2}{2\sigma_\pi^2}}~.
\end{eqnarray}
For the error $\sigma_\pi$, we averaged parallax errors of every bin. More details about the method can be found in \cite{zofi}.

\begin{figure*}
        \centering
        \subfloat[$\ang{0}<\phi<\ang{30}$]{
                \includegraphics[width=0.47\textwidth]{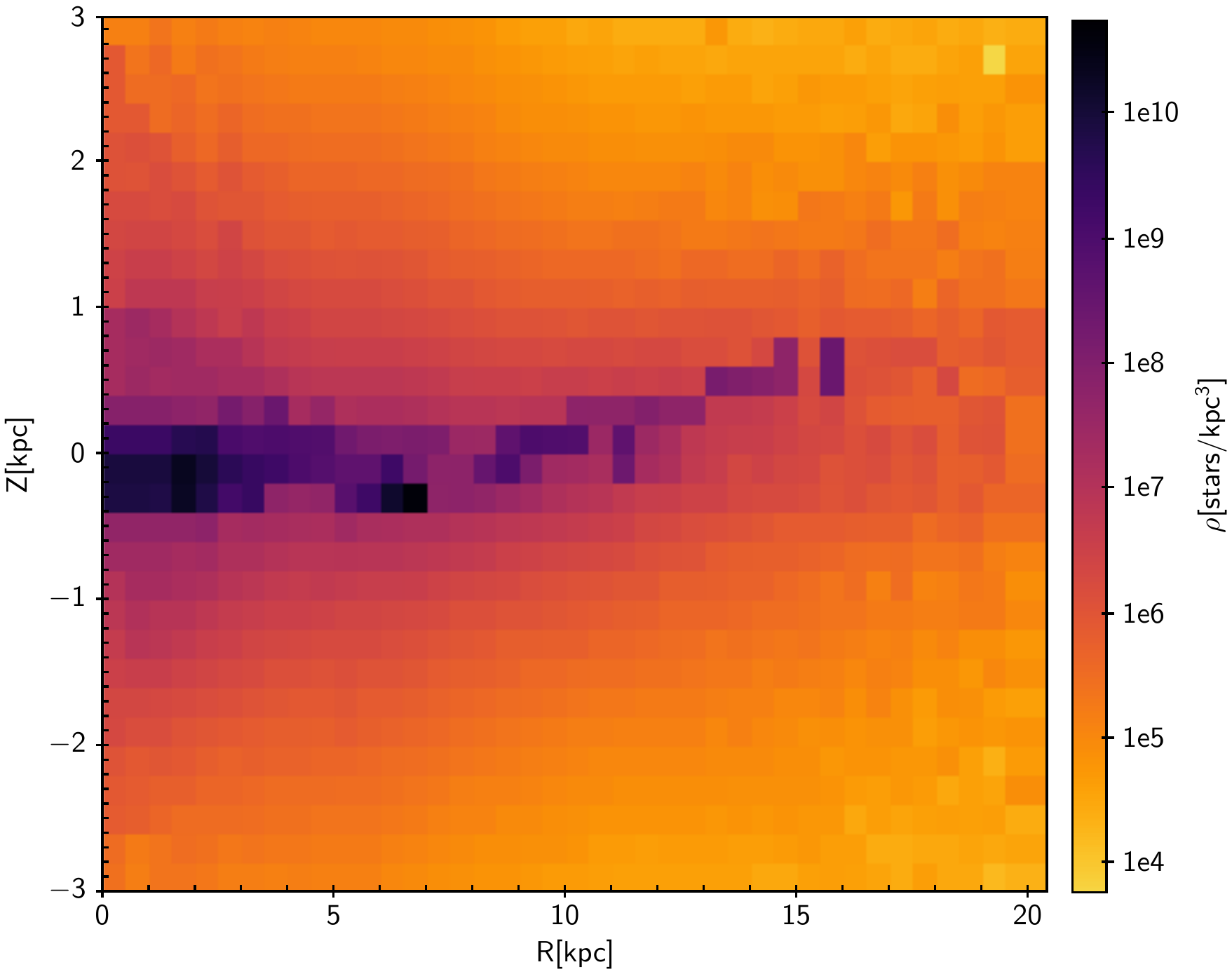}
        }
        \subfloat[$\ang{30}<\phi<\ang{60}$]{
                \includegraphics[width=0.47\textwidth]{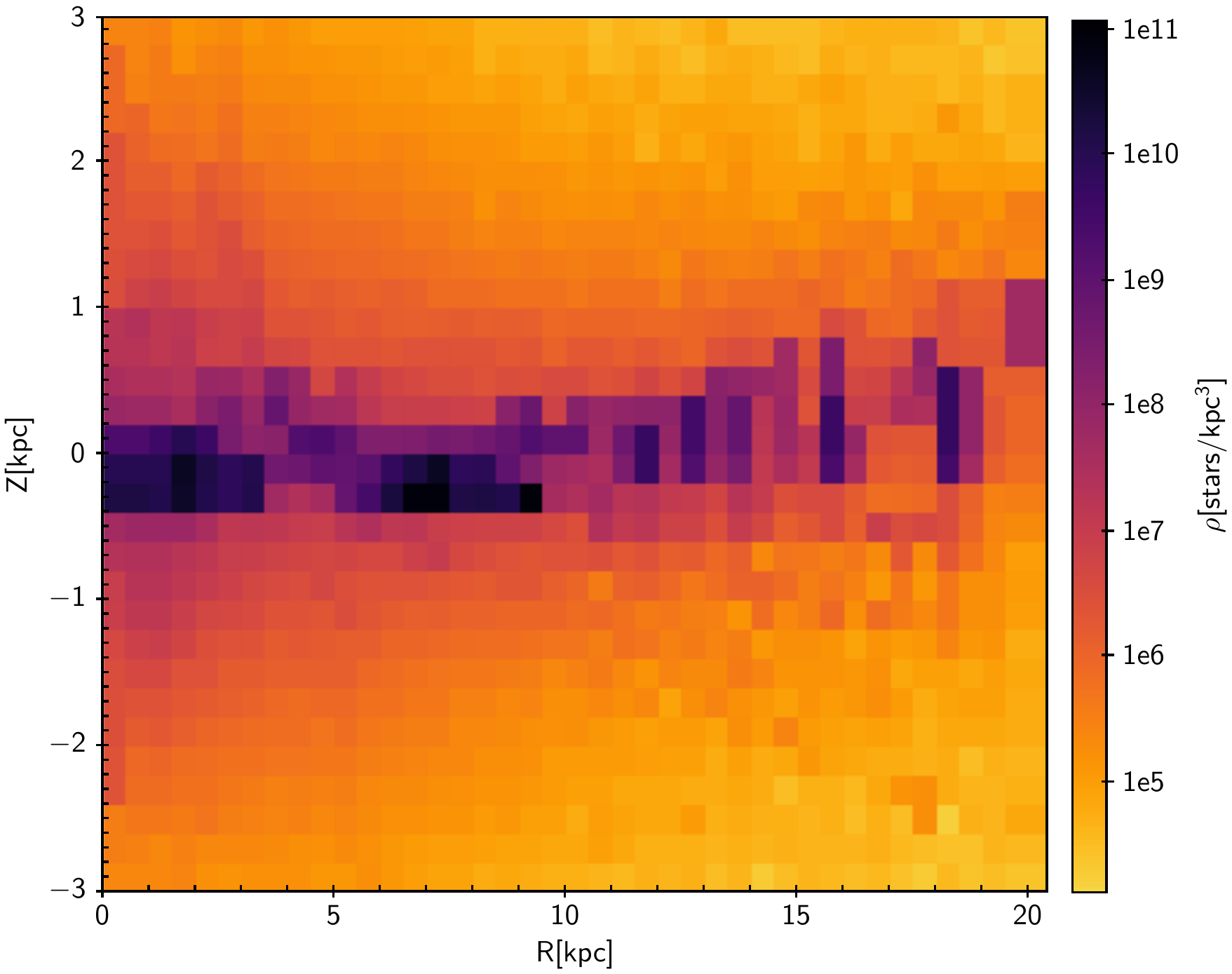}
        }
        \hspace{0mm}
        \subfloat[$\ang{60}<\phi<\ang{90}$]{
                \includegraphics[width=0.47\textwidth]{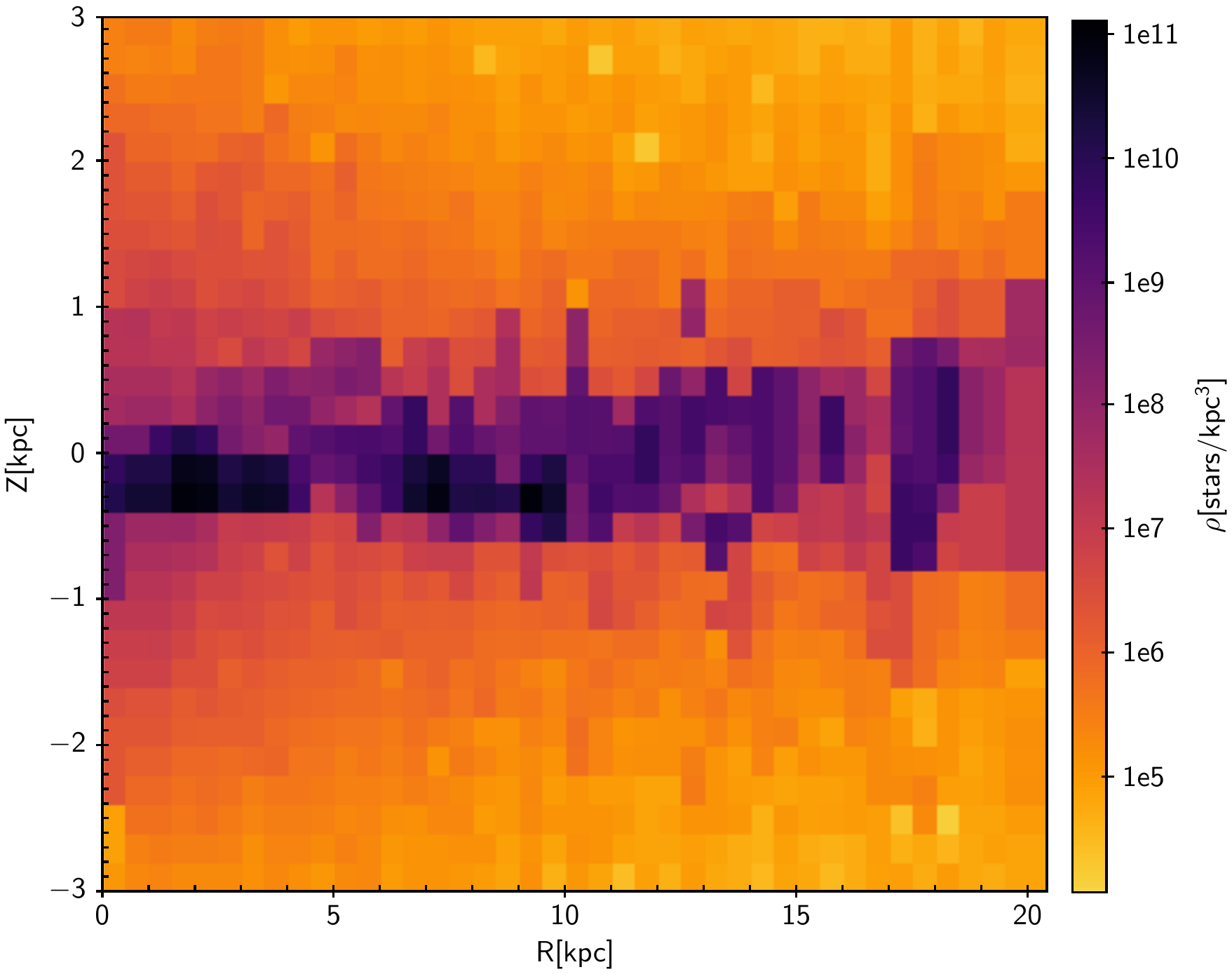}
        }
        \subfloat[$\ang{270}<\phi<\ang{300}$]{
                \includegraphics[width=0.47\textwidth]{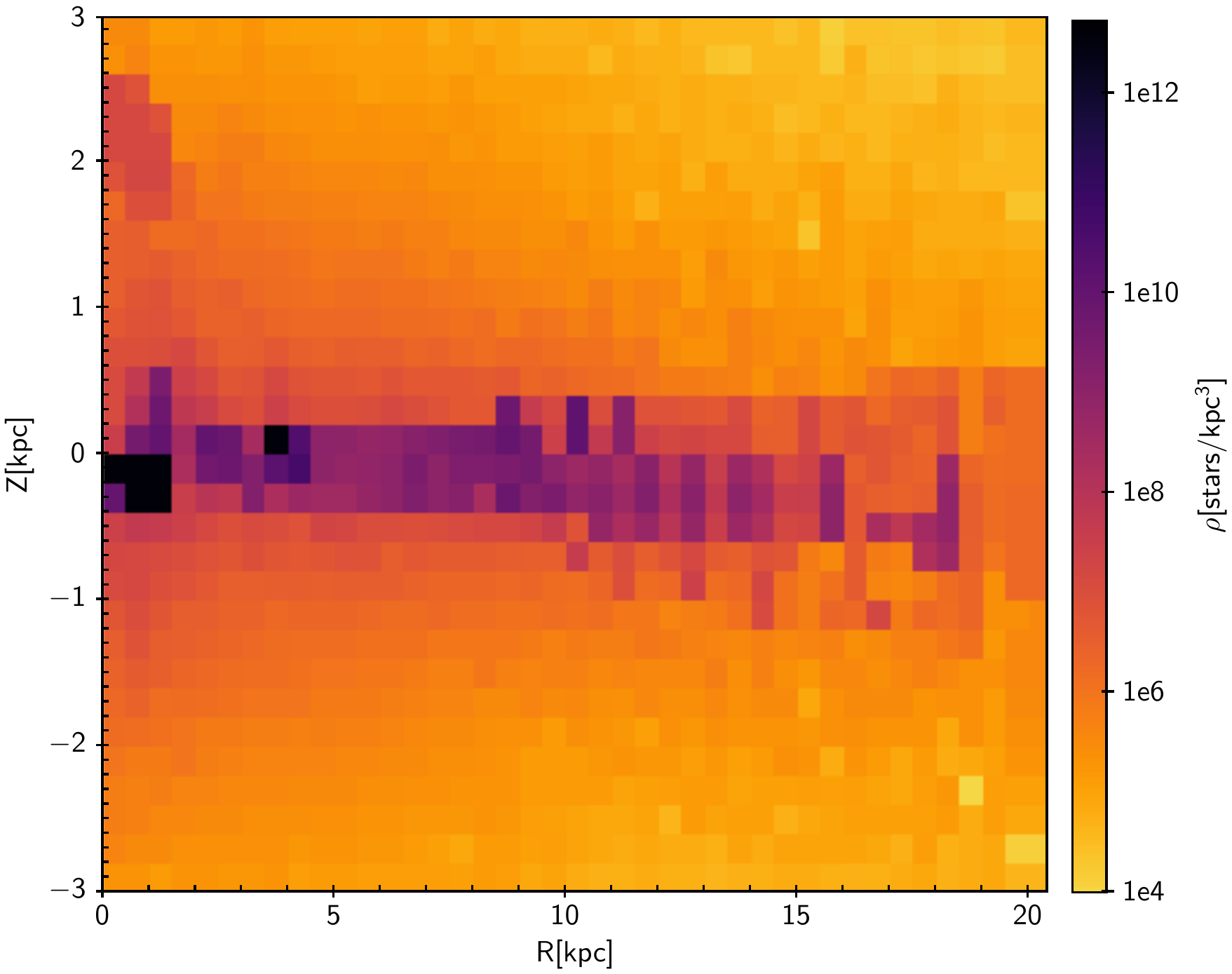}
        }
        \hspace{0mm}
        \subfloat[$\ang{300}<\phi<\ang{330}$]{
                \includegraphics[width=0.47\textwidth]{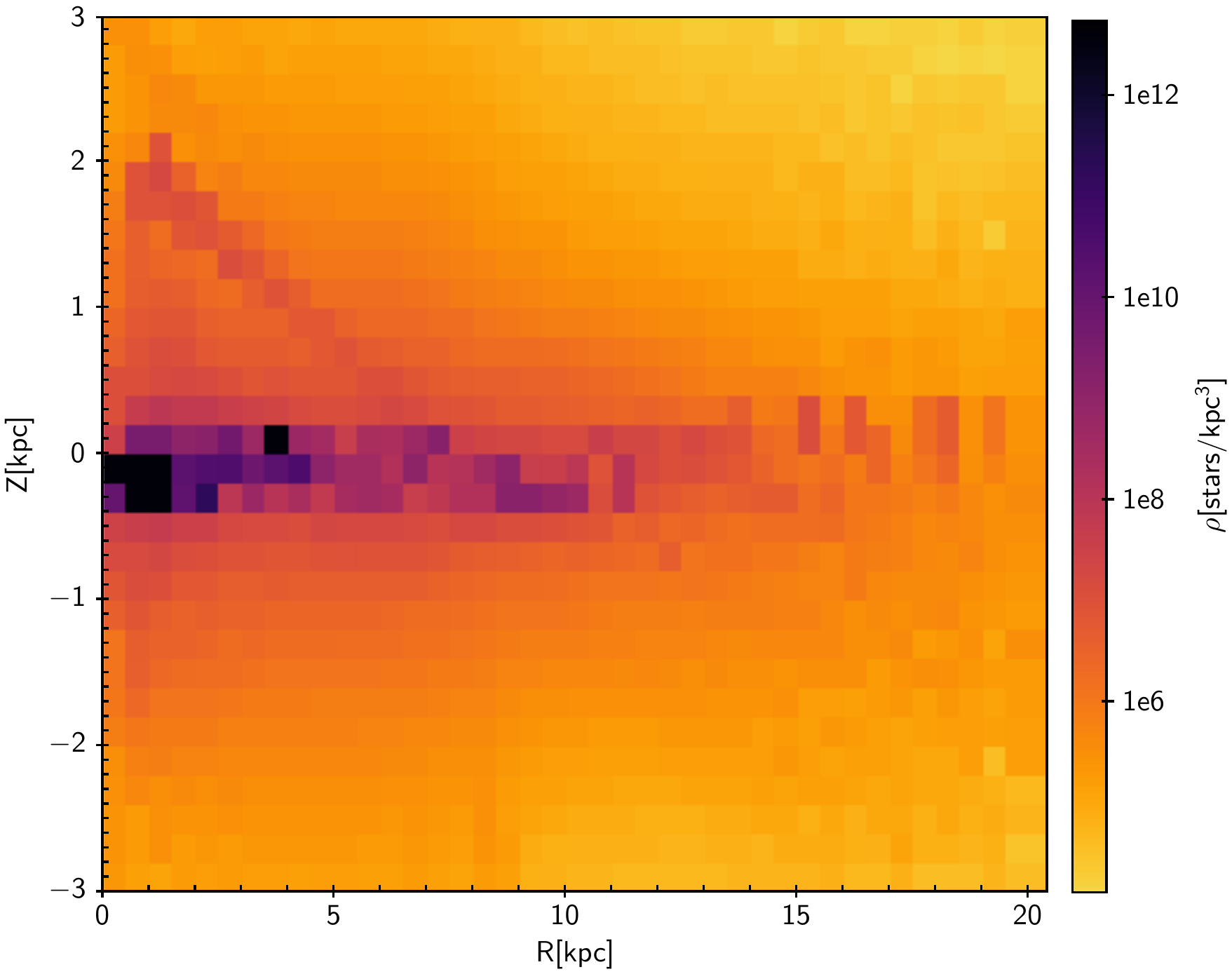}
        }
        \subfloat[$\ang{330}<\phi<\ang{360}$]{
                \includegraphics[width=0.47\textwidth]{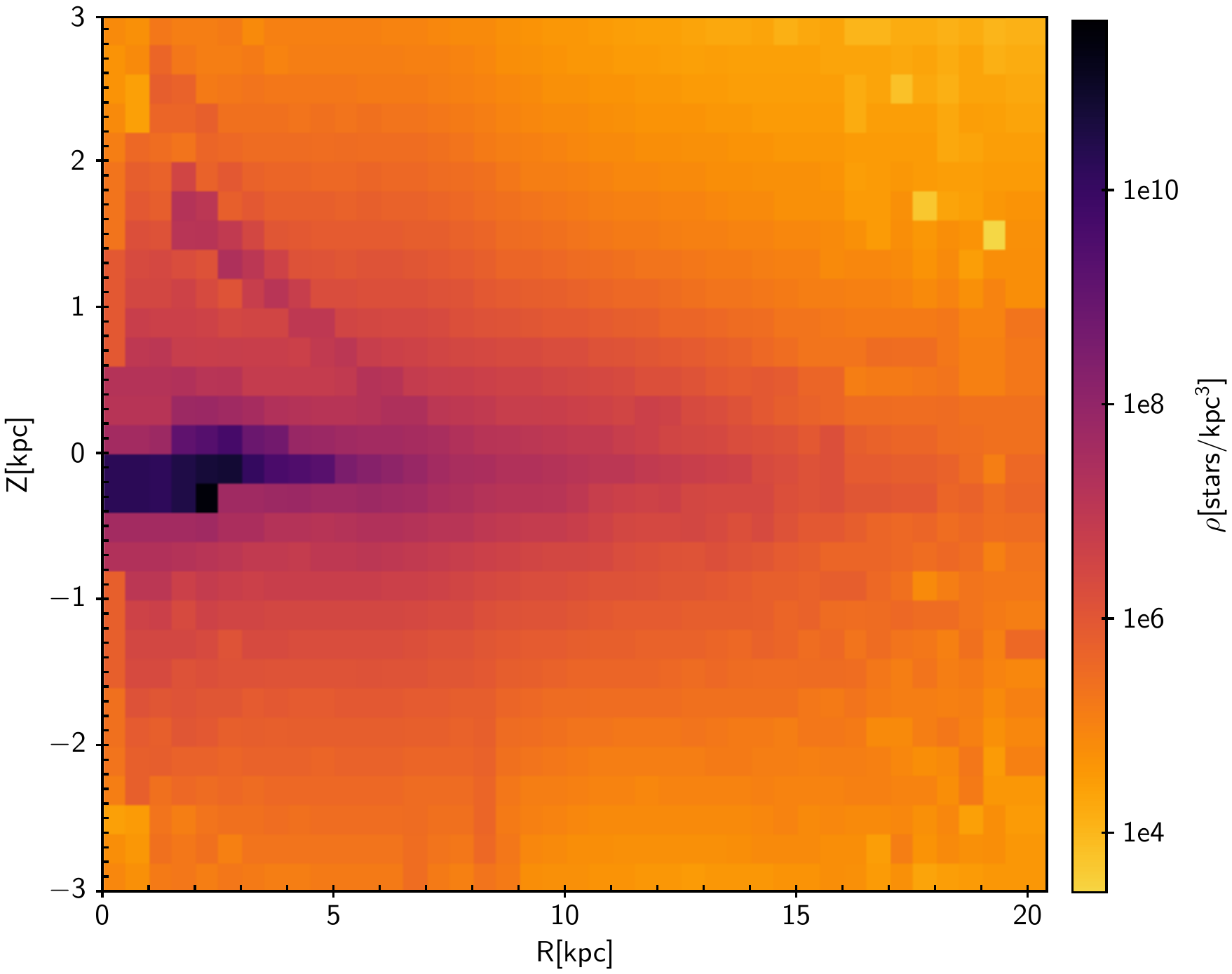}
        }
        \caption{Density maps in Galactocentric coordinates at various azimuths for the whole population (Sample 0).}\label{maps_edr3}
\end{figure*}

With the corrected star counts, we reveal density distribution up to 20 kpc, which can be seen in Fig. \ref{maps_edr3}. This result is almost identical with the one obtained with Gaia DR2 data \citep{zofi}. Similarly, we can see overdensities above the plane for azimuths between $\ang{300}$ and $\ang{360}$. As commented in \cite{zofi}, these structures are most likely a contamination, since they disappear after integrating the density through the whole disc.

We divided the data into bins of Galactic longitude $\ell$, Galactic latitude $b,$ and apparent magnitude $m$. For the values of $b$, we made bins of length $\ang{2}$ and corresponding $\ell$ in bins of $\ang{5}/cos(b)$. We divided each of the lines of sight in magnitude, binned with size $\Delta m=1.0$ between G=12 and G=19. We also made bins of $\Delta \pi=0.01$ mas in parallax. We did not use negative parallaxes because these affect the distribution of parallaxes and statistical properties. However, in our method we do not calculate the average distance from the average parallax. We used Lucy's method, which iterates the counts of the stars with positive parallaxes, until we obtained the final solution. This does not mean that we truncated the star counts with negative parallaxes; we simply did not use this information because it is not necessary with this approach.  Further details on this method and tests of its possible biases can be found in \cite{zofi}.

\section{Sample definition}

We analysed this density distribution to reveal the warp and the flare in the whole stellar population, which we will refer to as Sample 0. Moreover, we separated supergiants from the dataset in order to analyse their density distribution separately and find differences with the respect to the whole population. We used two different approaches to separate the supergiants, as described below.

\subsection{Sample 1}\label{sample1}
In the first approach, we chose only stars which we can be certain are supergiants. Based on the error of parallax, we calculated the interval of possible magnitudes $[M_{min},M_{max}]$ in the $G$- band for every star:

\begin{eqnarray}
M_{min}&=&m_{G}-5log_{10}(1/(\pi-\omega))-10-A_{G}(1/\pi)~ \nonumber \\
M_{max}&=&m_{G}-5log_{10}(1/(\pi+\omega))-10-A_{G}(1/\pi)~,
\end{eqnarray}
where $\pi$ is parallax and $\omega$ is parallax error. Then we chose only the stars with $-5<M_{min}<-10$ and $-5<M_{max}<-10$. In the end we had a dataset of $331546$ stars. 

As the distribution was not homogeneous, we divided stars in bins as follows. For $l<\ang{120}$ and $l>\ang{260}$ and $\lvert b \rvert < \ang{4}$, we had bins with $\Delta l=\ang{5}$ and $\Delta b=\ang{2}$. For the same range of $l$ for $\lvert b \rvert > \ang{4}$, we had $\Delta l=\ang{40}$ and $\Delta b=\ang{20}$. We binned the stars in $\ang{120}<l<\ang{260}$  in one bin. We also made bins $\Delta m=1.0$ in magnitude and $\Delta \pi=0.01$ mas in parallax.
In Fig. \ref{supergiants_distribution} (a), we show the distribution of the sources selected with this method.

\subsection{Sample 2}\label{sample2}
\begin{figure*}
        \centering
        \subfloat[]{
                \includegraphics[width=0.5\textwidth]{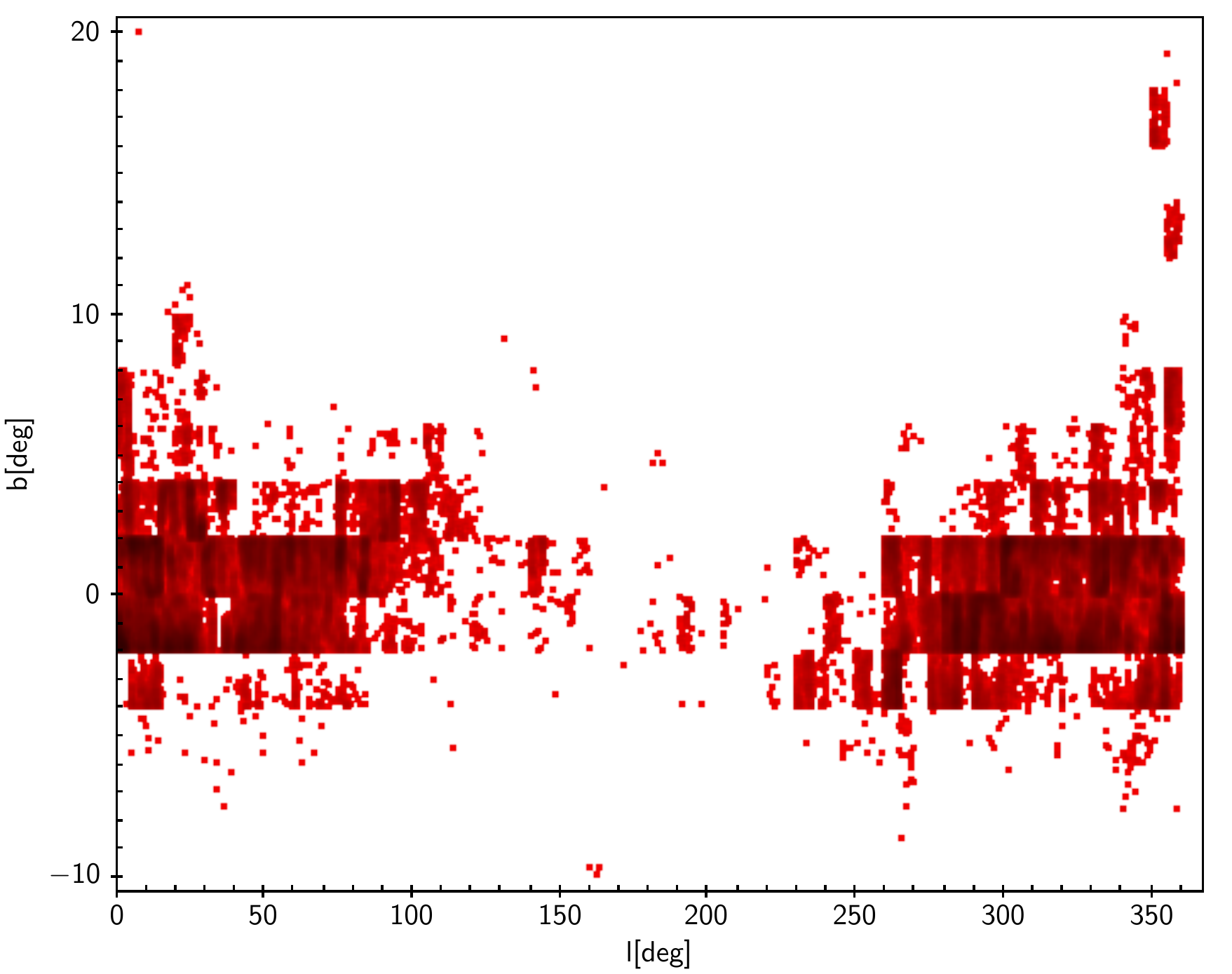}
        }
        \subfloat[]{
                \includegraphics[width=0.5\textwidth]{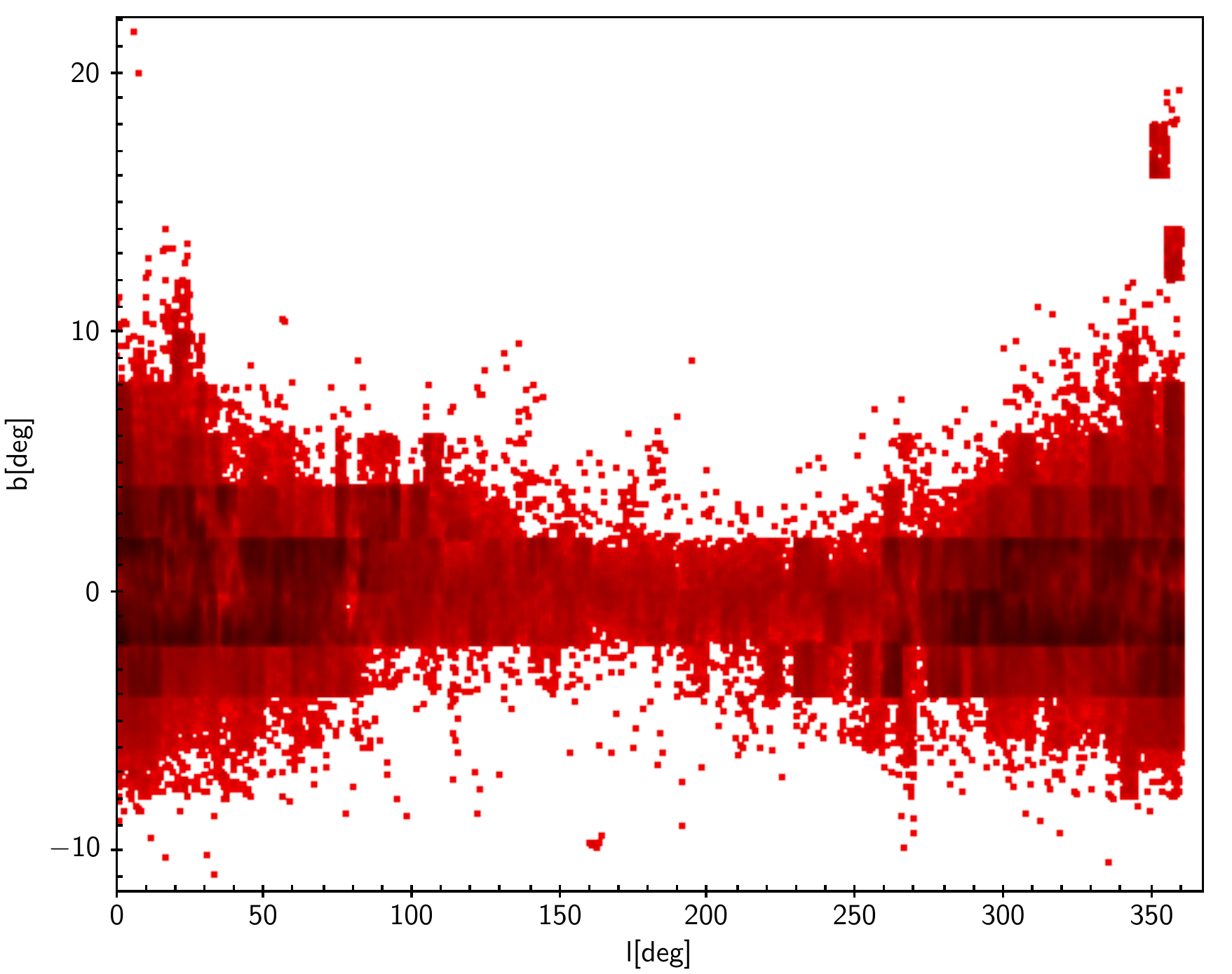}
        }
        \caption{Distribution of supergiants in galactic coordinates for $\lvert Z \rvert < 4$ kpc. Left: Sample 1. Right: Sample 2 (see Section \ref{sample1} and \ref{sample2} for details).}\label{supergiants_distribution}
\end{figure*}

\begin{figure}
        \centering
        \subfloat[]{
                \includegraphics[width=0.4\textwidth]{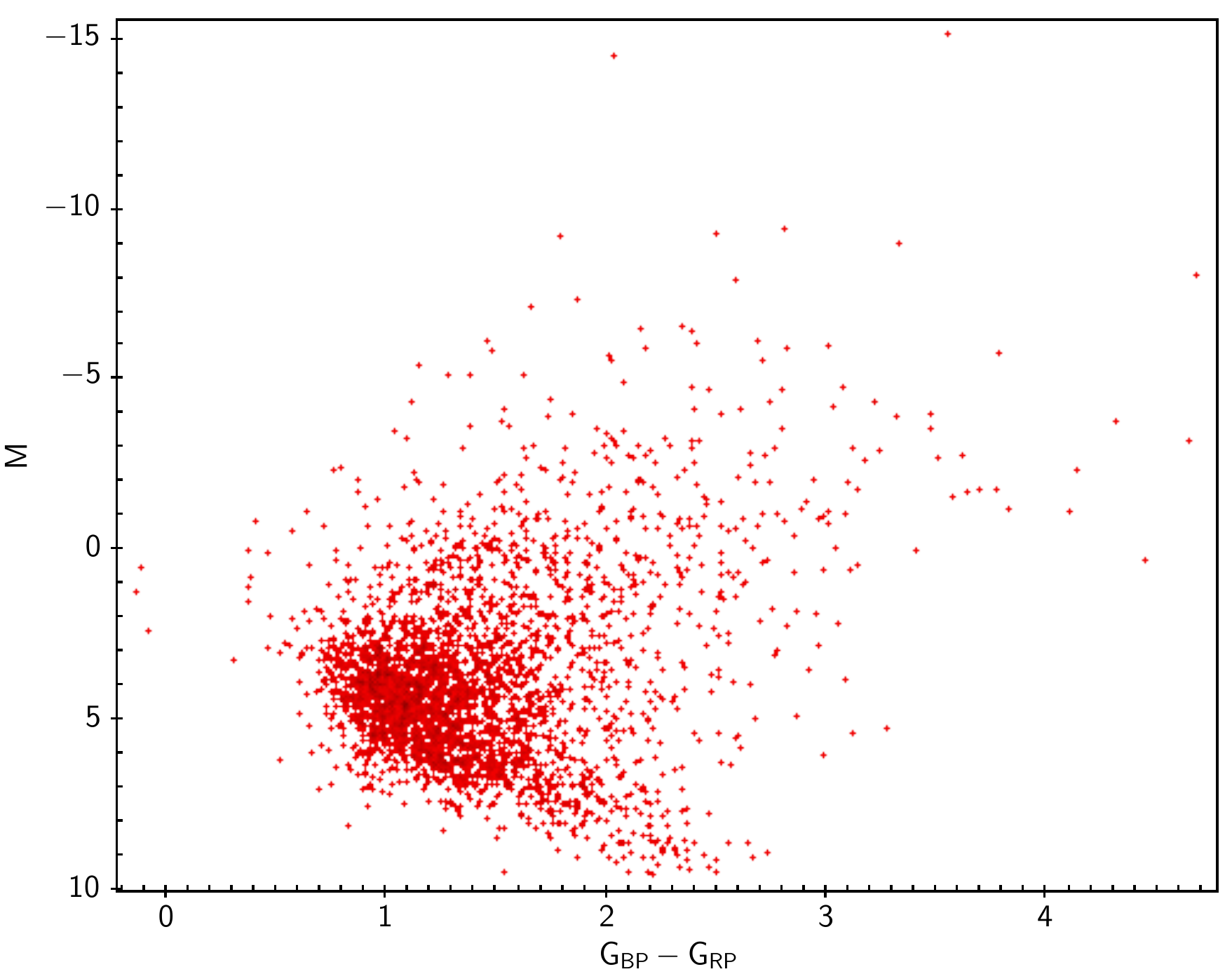}
        }
        \hspace{0mm}
        \subfloat[]{
                \includegraphics[width=0.4\textwidth]{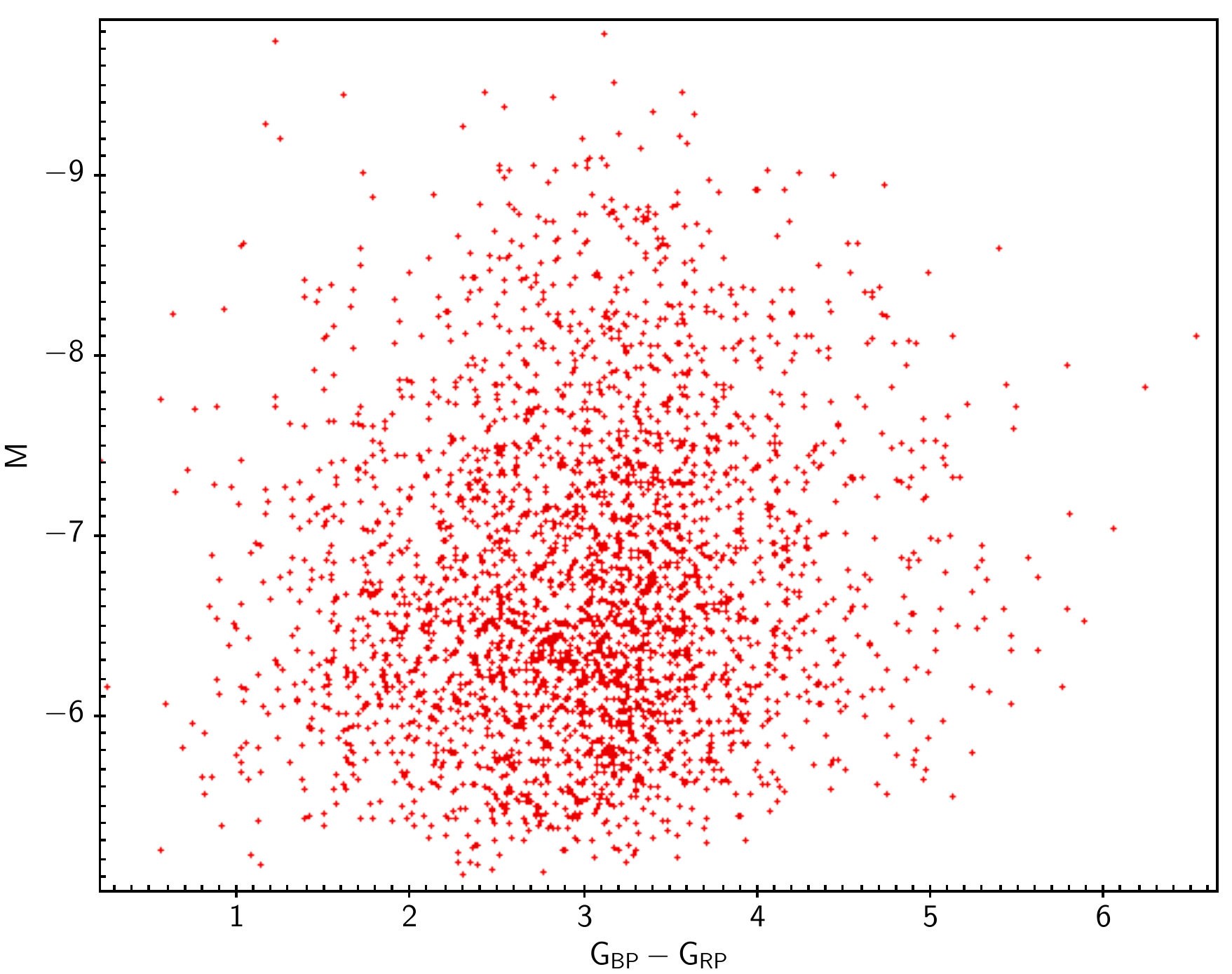}
        }
        \hspace{0mm}
        \subfloat[]{
                \includegraphics[width=0.4\textwidth]{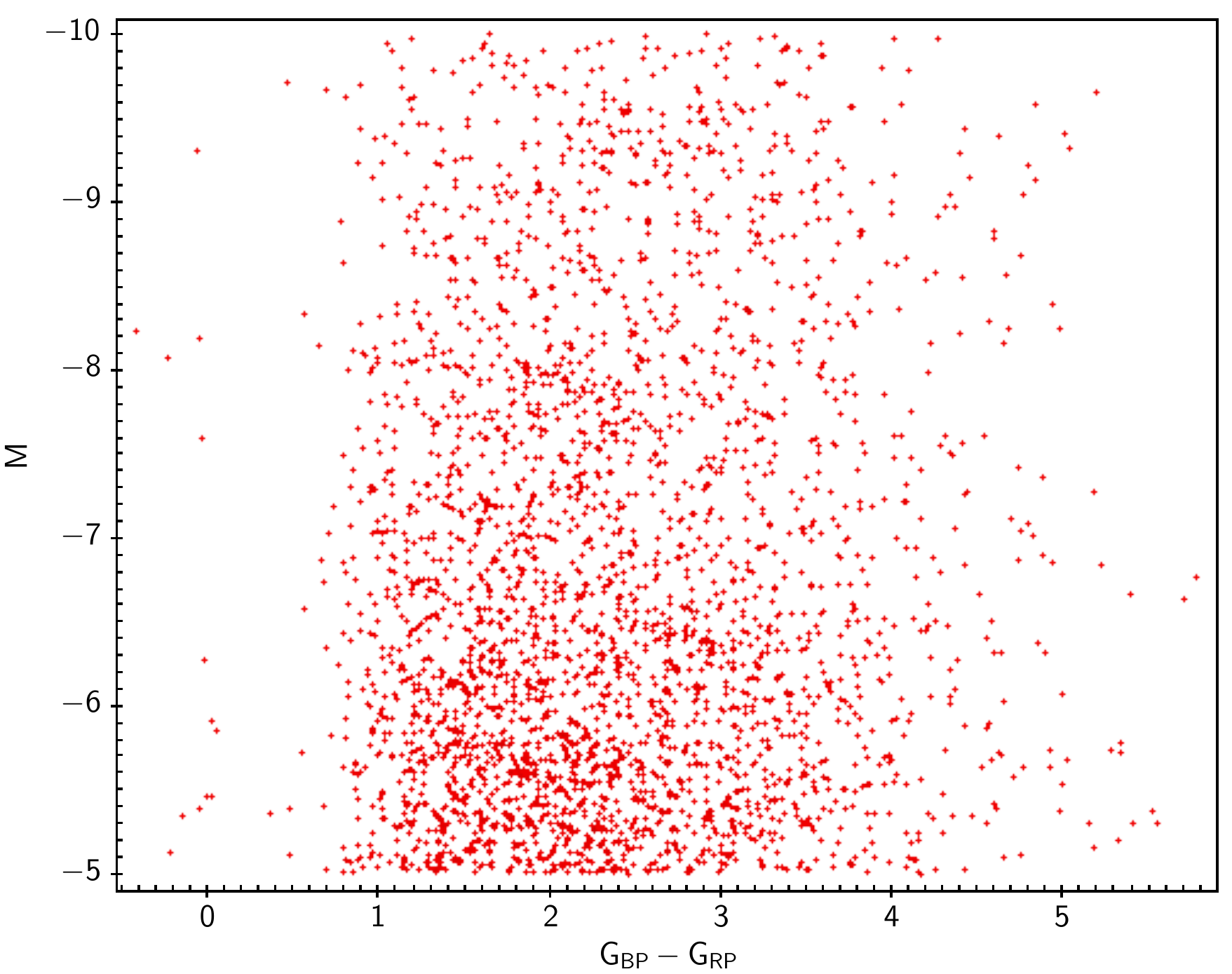}
        }
        \caption{Hertzsprung-Russell diagrams (HRD) for the three samples defined in the text. As the datasets are significantly large, in order to avoid the saturation of the diagram, we plot a randomly chosen sub-sample containing 3000 stars in each case. Top: Sample 0. Middle: Sample 1. Bottom: Sample 2.}\label{hr_diagrams}
\end{figure}

The second approach was less strict; we chose sources with absolute magnitude in the $G$- band within the interval $-5<M<-10$, regardless of the error of the magnitude. The advantage of this method is that we were complete, and therefore we could make statistical analysis of the data to investigate the flare. As this sample was distributed more homogeneously, we divided the stars in bins the same way as for the total population. In Fig. \ref{supergiants_distribution} (b), we show the distribution of the sources selected with this method. In the end, this sample contained about $1.9\cdot 10^{6}$ stars.
In Fig. \ref{hr_diagrams} we show the Hertzsprung-Russell diagrams (HRD) of all three of the samples.

\section{Warp analysis}\label{warp}
The first feature that we studied is the Galactic warp. We removed data for $\ang{90}<\phi<\ang{270}$ as in this part we did not have enough data. Following the approach of \cite{zofi}, we calculated the average elevation of the plane as 

\begin{equation}
z_w=\frac{\int_{z_{min}}^{z_{max}} \rho z \mathrm{d}z }{\int_{z_{min}}^{z_{max}} \rho \mathrm{d}z }
\end{equation}
Then, we fitted $z_w$ with Eq. (11) of \cite{zofi}, which represents a common warp model:

\begin{eqnarray}\label{9}
z_w=[C_wR(pc)^{\epsilon_w}sin(\phi-\phi_w)]~\mathrm{pc} ~,
\end{eqnarray}
where $C_w,\epsilon_w$ and $\phi_w$ are free parameters characterizing the warp. We did not account for the height of the Sun on the Galactic plane because some recent studies \citep[e.g.][]{cheng_preces} suggest that warp starts at smaller radii than previously thought and, therefore, the Solar neighbourhood could be slightly warped. For Sample 0, we find values of warp parameters:

\begin{eqnarray}
c_w &=& 1.42\pm0.15\times10^{-8} ~\mathrm{pc} ~, \nonumber \\
\epsilon_w&=& 2.43 \pm 0.65~, \\
\phi_w&=& -9.77 \pm 7.23 \fdg ~, \nonumber
\end{eqnarray} 
while for Sample 1, we find:
\begin{eqnarray}
c_w &=& 1.92 \pm 0.08 \times 10^{-4} ~\mathrm{pc}~, \nonumber \\
\epsilon_w&=& 1.54 \pm 0.18~, \\
\phi_w&=& -8.23 \pm 2.95 \fdg ~ \nonumber
\end{eqnarray}
and for Sample 2:
\begin{eqnarray}
c_w &=& 4.85 \pm 0.18 \times 10^{-5} ~\mathrm{pc}~, \nonumber \\
\epsilon_w&=& 1.66 \pm 0.17~, \\
\phi_w&=& -0.73 \pm 2.55 \fdg ~. \nonumber
\end{eqnarray} 
The error of $c_w$ stands for the error of the amplitude alone, without the variations of $\epsilon_w$ and $\phi_w$. For the fit, we used the function \textit{curve fit} from the python \textit{SciPy} package, which uses non-linear least squares to fit a function to data.

\begin{figure}  
        \includegraphics[width=9cm]{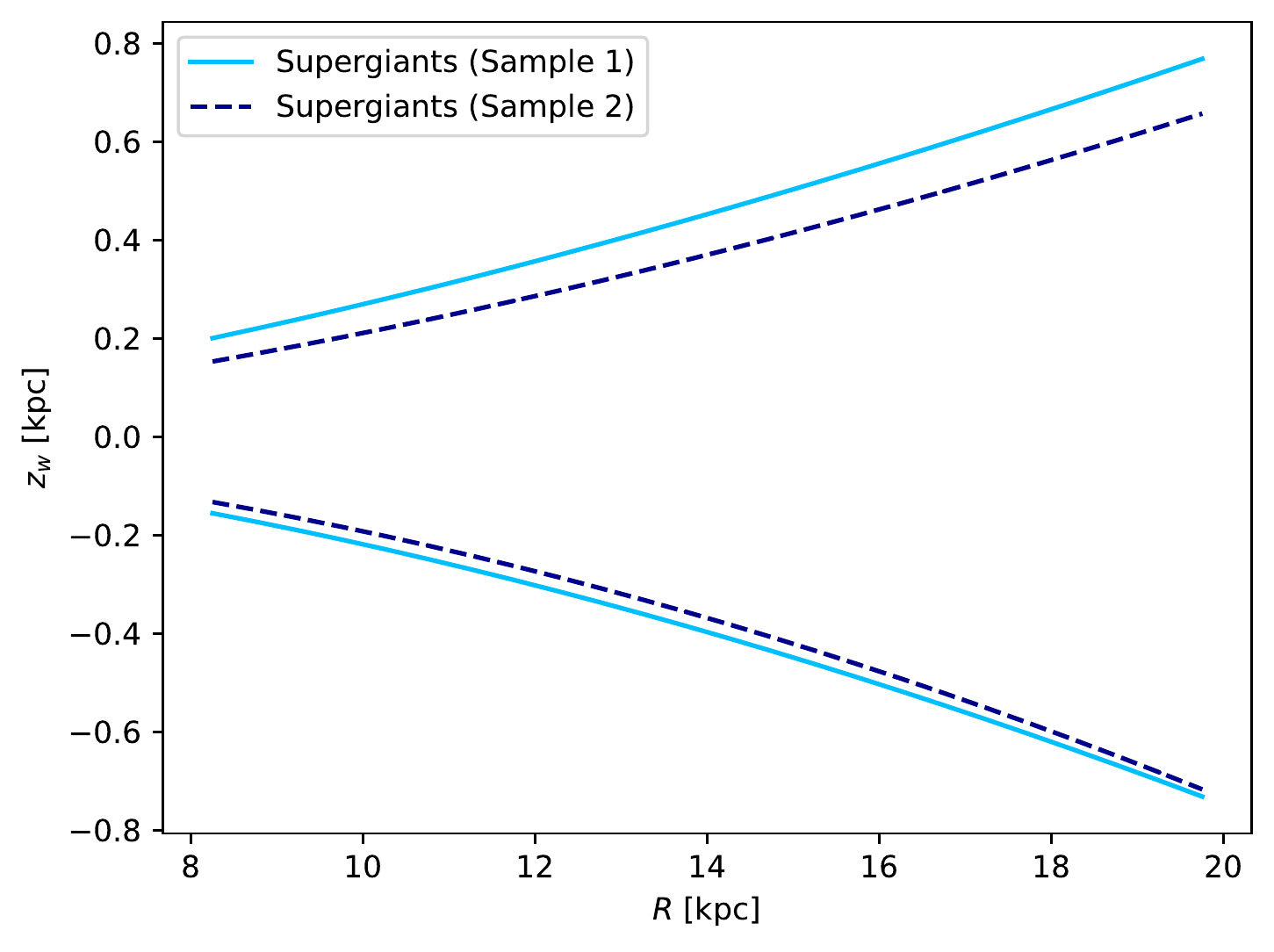}
        \caption{Comparison of fits of minimum and maximum warp amplitudes for supergiants, chosen by two different approaches (see Sections \ref{sample1} and \ref{sample2} for details).}\label{supergiants_warp}
\end{figure}

\begin{figure}
        \includegraphics[width=9cm]{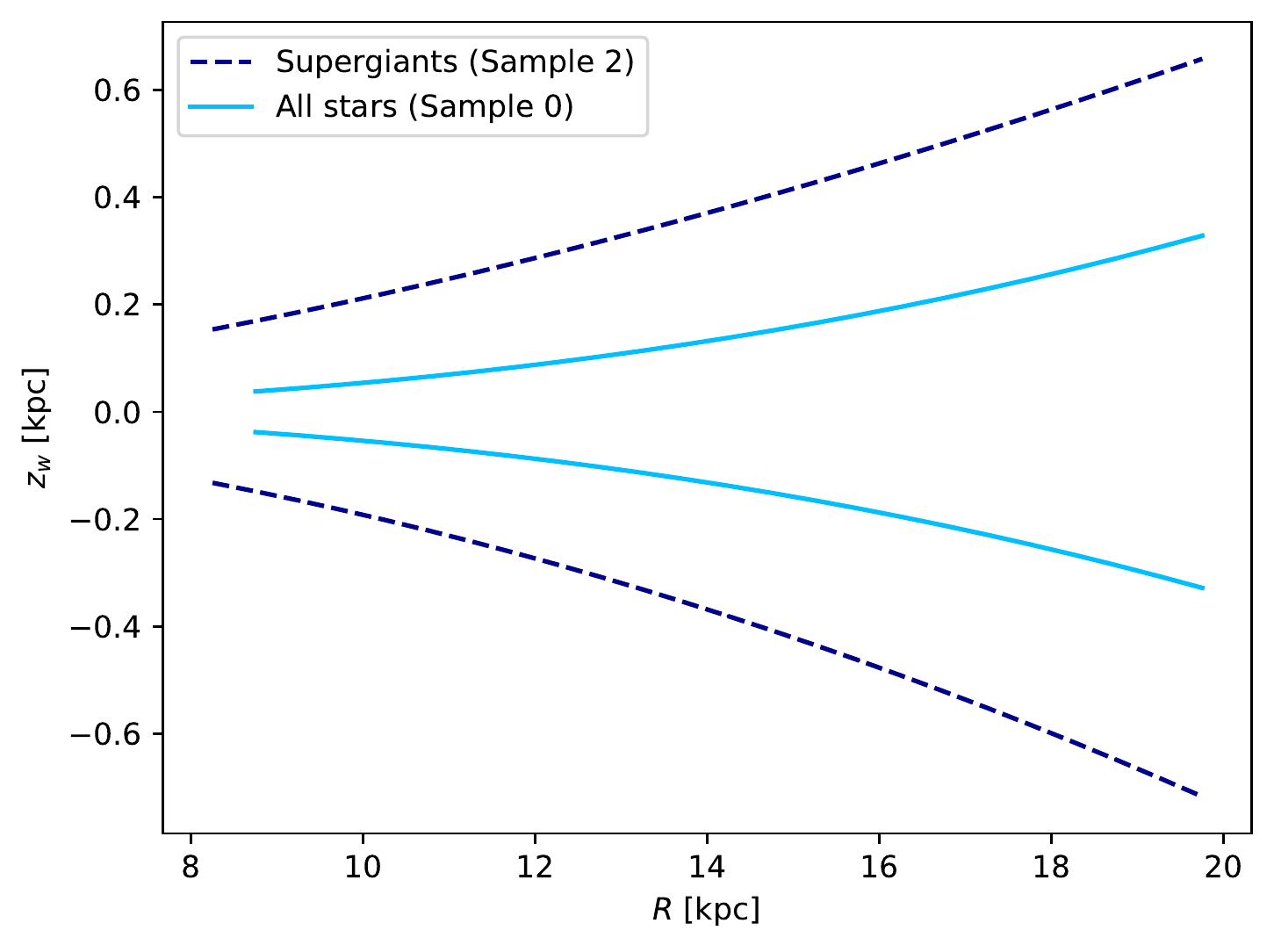}
        \caption{Comparison of fits of maximum and minimum warp amplitudes for the whole population (Sample 0) and supergiants (Sample 2).}\label{warp_superg_whole_comparison}
\end{figure}

In Fig. \ref{supergiants_warp} we compare the warp amplitude for supergiants for both samples (Sample 1 and Sample 2). The two samples are in very good agreement, yielding warp amplitude with only negligible differences, showing that the Sample 2 is not significantly contaminated. 
In Fig. \ref{warp_superg_whole_comparison}, we plot the maximum and minimum amplitudes of warp for the whole population (Sample 0), compared with the supergiants (Sample 2). We confirm that the warp amplitude for the whole population is almost identical to what we found with Gaia DR2 data \citep{zofi}. We obtain a maximum warp amplitude of $z_w=0.360$ kpc and minimum of $z_w=-0.375$ kpc at a distance of $R=[19.5,20]$ kpc, which is slightly higher than the result obtained with Gaia DR2 data, with a small asymmetry between the northern and the southern warp. From comparison with the supergiants, it is clear that the warp amplitude of supergiants is significantly larger that of the whole population, reaching an amplitude twice as large at a distance of $R=[19.5,20]$ kpc. As supergiants are a young population, a few tens of Myr old on average \citep[e.g.][]{superg_vek}, whereas the whole population is $\sim 6-7$ Gyr old \citep[e.g.][]{disc_vek}, there is a clear relationship between the warp amplitude and the age of the studied population, thus confirming conclusions of previous studies \citep{zofi,haifeng_2020}. 


\section{Flare analysis}
In order to investigate the Galactic flare, we considered the density distribution of the Galactic disc as consisting of a thick and a thin component. We adopted the model of the flared disc presented by \cite{martin_flare} in the form:

\begin{eqnarray}\label{discmodel}
\rho_{disc}(R,z)&=&\rho_{thin}(R,z) +\rho_{thick}(R,z), \nonumber \\
\rho_{thin}(R,z)&=&(1-f)~\rho_\odot ~ \mathrm{exp}\left(\frac{R_\odot}{h_r}+\frac{h_{r,hole}}{R_\odot}\right) \nonumber \\
&\times& \mathrm{exp}\left(-\frac{R}{h_r}-\frac{h_{r,hole}}{R}\right)\mathrm{exp}\left(-\frac{\lvert z \rvert}{h_{z,thin}}\right),  \\
\rho_{thick}(R,z)&=&f~\rho_\odot~\mathrm{exp}\left(\frac{R_\odot}{h_r}+\frac{h_{r,hole}}{R_\odot}\right) \nonumber \\
&\times& \mathrm{exp}\left(-\frac{R}{h_r}-\frac{h_{r,hole}}{R}\right)\mathrm{exp}\left(-\frac{\lvert z \rvert}{h_{z,thick}}\right), \nonumber \\
\nonumber
\end{eqnarray}
where the Galactocentric cylindrical coordinate system ($R,z$) is used. This model takes into account the thick and the thin discs, with an exponential decrease in density in the horizontal and the vertical directions. $h_r$ is the scale length of the whole disc. $h_{z,thin}$ and $h_{z,thick}$ are the scale heights of the thin and the thick discs, respectively. The deficit of stars in the inner region of the disc is characterised by the $h_{r,hole}$ parameter \citep{corr04}. Since we focus on the remote regions of the Galaxy, we kept $h_{r,hole}$ constant at $h_{r,hole}=3.74$ kpc \citep{martin_flare}. The Galactocentric distance of the Sun is $R_\odot=8.25$ kpc and $\rho_\odot$ is the volume mass density of the Galactic disc in the solar neighbourhood. The $f$ parameter represents the ratio of thick and thin stars in the solar neighbourhood and we kept it at $f=0.09$. We experimented with setting $f$ as a free parameter, but it was varying only slightly, with negligible influence on $h_z$, and therefore we kept it at the local value as done by \cite{martin_flare}. The parameters $\rho_\odot$, $h_{r}$, $h_{z,thin}$, and $h_{z,thick}$ are free and their values were then determined from the fitting procedure. The thin and thick discs were divided based on the geometry of the density profiles in the vertical plane.

\begin{table*}
	\centering
	\begin{tabular}{|c c c c|}
		\hline
		Reference & Data source & Constrains & Scale length [kpc] \\
		\hline\hline
		This work & Gaia EDR3 & $\phi \in \left[330^{\circ}, 30^{\circ}\right]$; the whole population & $2.19 \pm 0.18$ \\
		This work & Gaia EDR3 & $\phi \in \left[330^{\circ}, 30^{\circ}\right]$; supergiants (Sample 2) & $1.99 \pm 0.13$ \\
		\cite{zofi} & Gaia DR2 & the whole population & $2.29 \pm 0.08$ \\
		\cite{li} & Gaia DR2 & OB stars & $2.10 \pm 0.01$ \\
		\cite{haifeng} & LAMOST & RGB, thin disc, $R \leq 11$ kpc & $2.13 \pm 0.23$ \\
		\cite{haifeng} & LAMOST & RGB, thick disc, $R \leq 11$ kpc & $2.72 \pm 0.57$ \\
		\cite{martin_flare} & SDSS-SEGUE & thin disc & $2.0^{+0.3}_{-0.4}$ \\
		\cite{martin_flare} & SDSS-SEGUE & thick disc & $2.5^{+1.2}_{-0.3}$ \\[1ex] 
		\hline
	\end{tabular}
	\caption{Scale length of the Galactic disc fitted by Eq. (\ref{equ}) for the whole population and supergiants (Sample 2), compared with other works.}
	\label{tab-comp1}
\end{table*}

\subsection{Method}\label{sec-meth}
To find the horizontal and vertical star distribution in the Galactic disc, we used the density maps (Fig. \ref{maps_edr3}). We carried the fitting procedure out in two steps: 1) investigating the density profile in the Galactic equatorial plane; 2) investigating the density profile in the vertical direction.

\subsection{Density profile in the Galactic equatorial plane}\label{sec-plane}
\begin{figure}  
        \includegraphics[width=9cm]{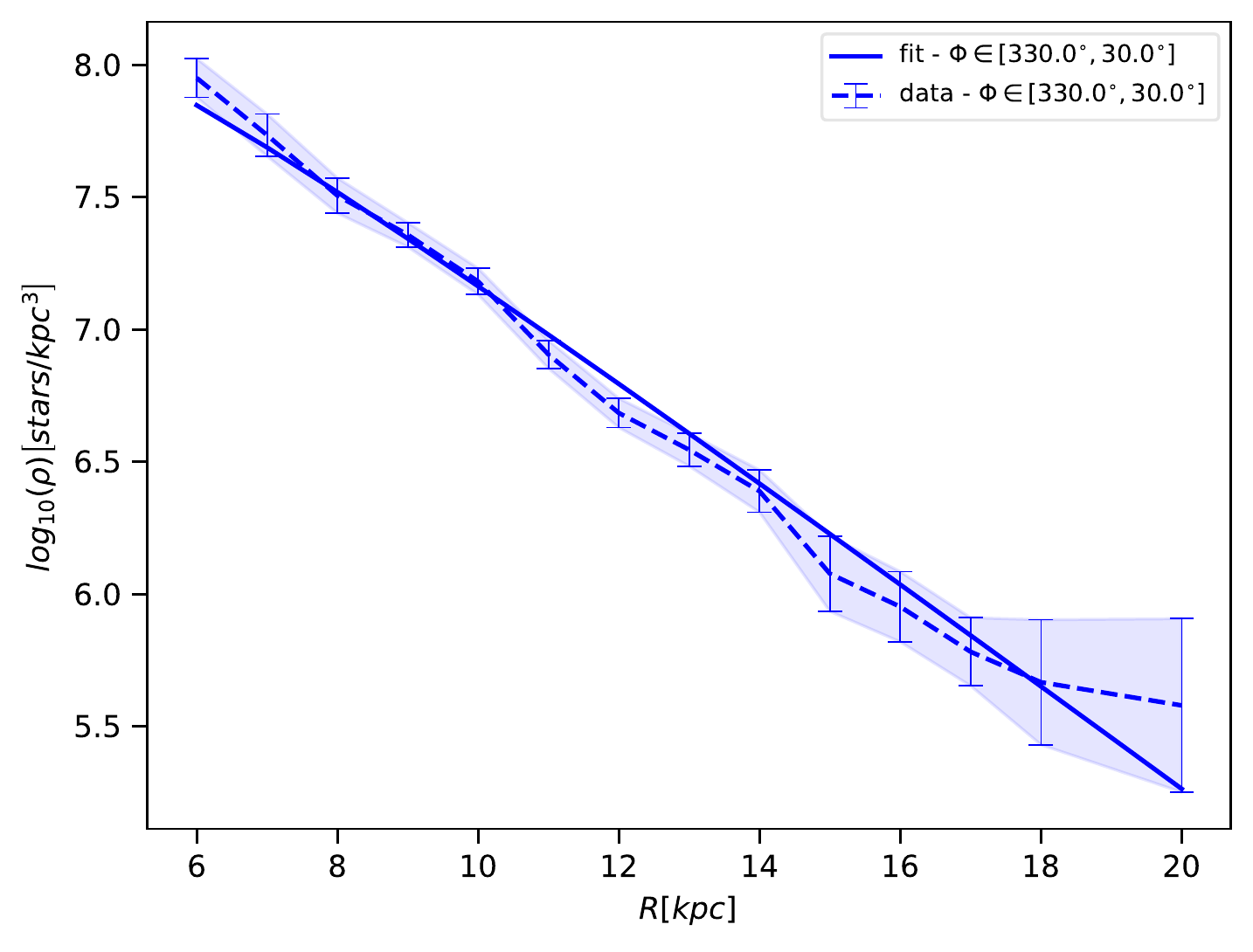}
        \caption{Dependence of the density on the Galactocentric distance in the Galactic equatorial plane for the azimuth $\phi \in \left[330^{\circ}, 30^{\circ}\right]$. The data points were obtained as weighted mean in bins of size $1$ kpc in $R$ and $0.4$ kpc in $\lvert z|,$ and were fitted with the model defined in Eq. (\ref{equ}).}\label{dens60}
\end{figure}

First of all, we focused on the Galactic plane in order to fit the scale length of the Galactic disc. We applied the following constrains on the data: we used stars with $|z|<0.2$ kpc and Galactocentric distances $R \in \left[5,20\right]$ kpc; the bin size was $0.4$ kpc in $z$ and $1$ kpc in $R;$ and we only considered bins with number of stars $N \geq 50$. We applied the disc model with exponential decrease in $R$ in the following form: 

\begin{eqnarray}\label{equ}
\rho(R)&=&\rho_\odot ~ \mathrm{exp}\left(\frac{R_\odot}{h_r}+\frac{h_{r,hole}}{R_\odot}\right) \nonumber\\ &\times& \mathrm{exp}\left(-\frac{R}{h_r}-\frac{h_{r,hole}}{R}\right) ~,
\end{eqnarray}
with $h_r$ and $\rho_{\odot}$ as free parameters to be fitted. We defined the azimuthal angle $\phi$ to be measured from the centre-Sun-anticentre direction towards the Galactic rotation, going from \ang{0} to \ang{360}. Since we could not distinguish the individual populations, we neglected the contribution of the thick disc stars in the Galactic equatorial plane. We applied the weighted minimum chi-square method on the data to obtain the values of the fitting parameters in the plane of the Galactic disc given by Eq.(\ref{equ}). The best results of the fitting procedure in the equatorial plane are presented in Table \ref{tab-comp1}, where the comparison with other works is also included. The density profiles in the Galactic plane for the whole data sample are plotted in Fig. \ref{dens60}. Comparing the scale length for various azimuths, one can see the slight dependence of the $h_r$ on $\phi$, especially for larger Galactic azimuths; $1.71\pm0.18$ kpc and $1.58\pm0.15$ kpc for $\Phi \in \left[300^\circ, 330^\circ\right]$ and for $\Phi \in \left[30^\circ, 60^\circ\right]$, respectively. On the other hand, variations of the scale length near the centre-Sun-anticentre direction is not present ($2.04\pm0.05$ kpc and $2.26\pm0.07$ kpc for $\Phi \in \left[330^\circ, 0^\circ\right]$ and for $\Phi \in \left[0^\circ, 30^\circ\right] $, respectively), and the average value of the scale length $h_r = 2.19 \pm0.08$ kpc (for $\Phi \in \left[330^\circ, 30^\circ\right]$) with little dependence on azimuth. The results are in agreement with previous works. For example, \cite{li} present $h_r=2.10\pm0.01$ kpc for OB stars using Gaia DR2 data, and \cite{zofi} show  $h_r=2.29\pm0.08$ kpc using Gaia DR2 data. On the other hand, \cite{yu} show $h_r = 1.17 \pm 0.05$ kpc for OB stars.

\begin{figure}  
        \includegraphics[width=9cm]{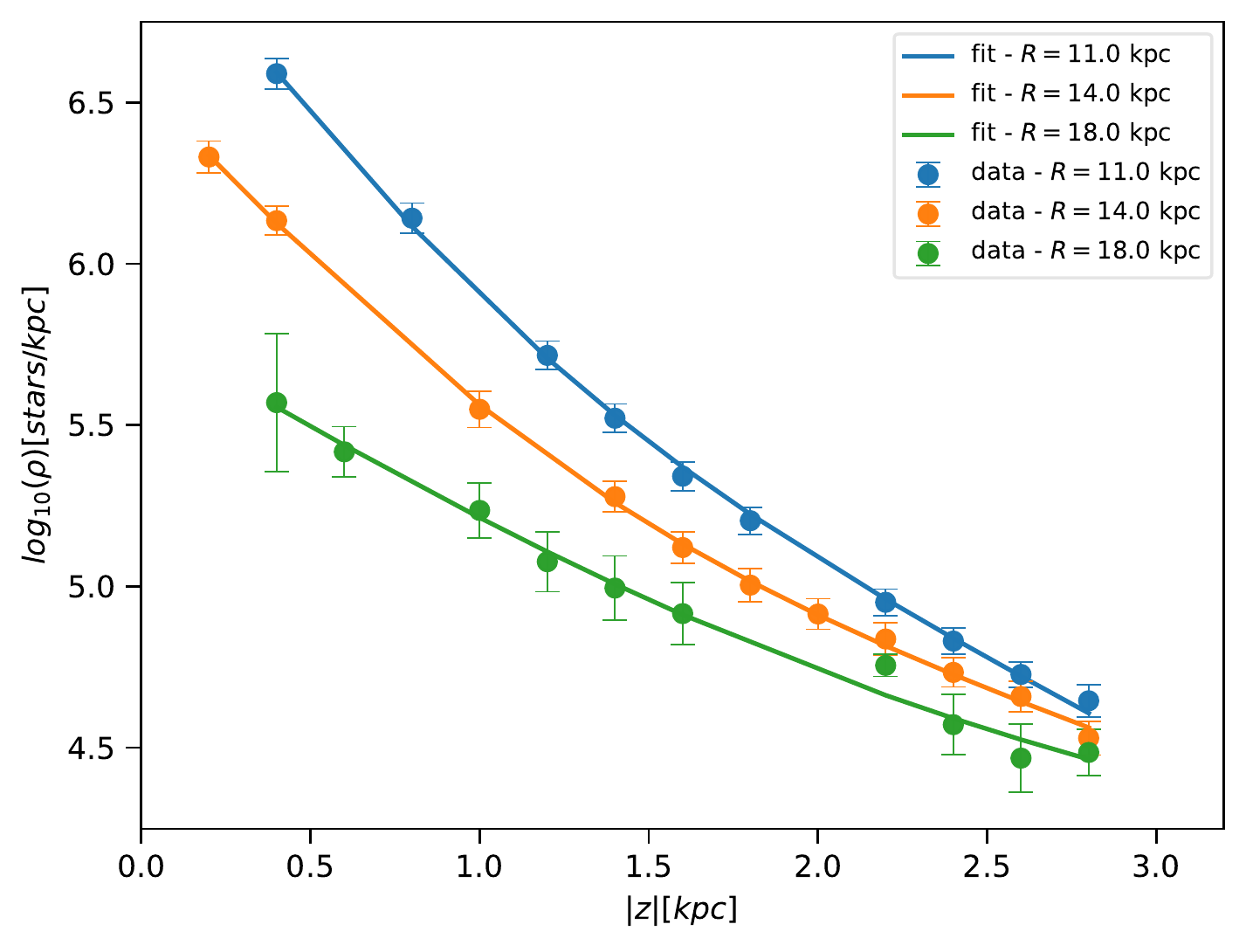}
        \caption{Dependence of the density on $|z|$ for various values of Galactocentric distance. The Galactic azimuth is $\phi \in \left[330^{\circ}, 30^{\circ}\right]$. The data points were obtained as weighted mean in bins of size $1$ kpc in $R$ and $0.2$ kpc in $|z|$.}\label{zprof}
\end{figure}

\subsection{The scale height} \label{sec-vertical}

\begin{figure}  
        \includegraphics[width=9cm]{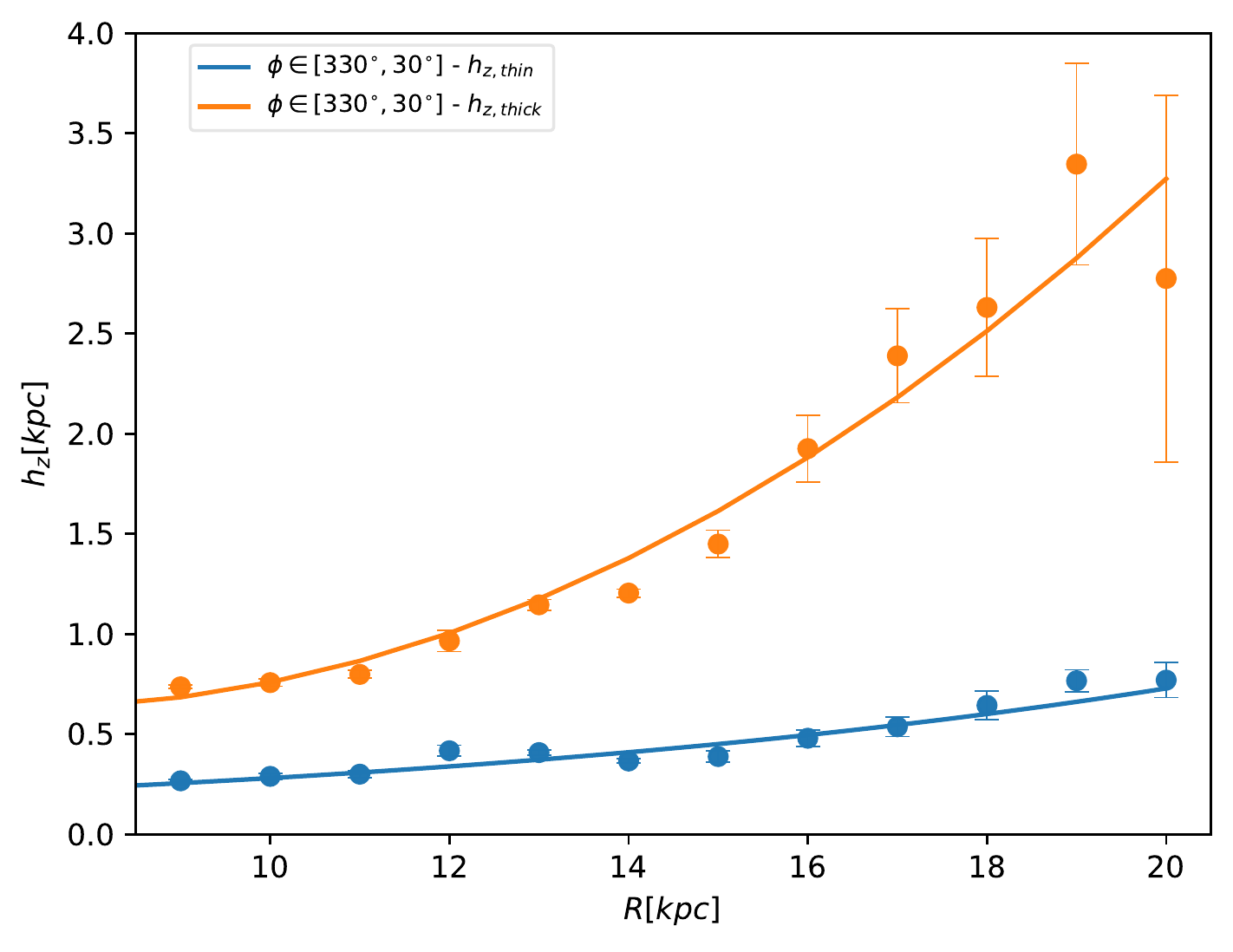}
        \caption{Dependence of the scale height of the thick and the thin discs on the Galactocentric distance. The Galactic azimuth is $\phi \in \left[330^{\circ}, 30^{\circ}\right]$. The dashed line is the second-order polynomial fit to the data points.}\label{hz}
\end{figure}

In order to investigate the flaring of the Galactic disc, we fitted the vertical density profiles of the whole data sample, which are presented in Fig. \ref{maps_edr3}, with the model of flared thick and thin discs described by Eq. (\ref{discmodel}). We applied the weighted minimum chi-square method to obtain the values of the scale height of the thick and thin discs, while the scale length calculated in Sec. \ref{sec-plane} remained fixed. We divided the data in bins with size $\Delta R = 1$ kpc and $\Delta z = 0.2$ kpc. The vertical density profiles for various values of Galactocentric distances ($R=11$ kpc, $R=14$ kpc and $R=18$ kpc) are plotted in Fig. \ref{zprof}. The values of the scale height are presented in Table \ref{tab-comp2} and shown in Fig. \ref{hz}. The $h_z$ fitting function is a polynomial of the second order. Here, we stress again that we divide the Galactic disc into thin and thick discs, based solely on geometric properties.

\begin{table}
        \centering
        {\small
        	\begin{tabular}{|c c c|}
        		\hline
        		$R$ [kpc] & $h_{z,thin}$ [kpc] & $h_{z,thick}$ [kpc] \\	
        		\hline\hline
                5.0 & $0.10 \pm 0.01$ & $0.50 \pm 0.02$ \\ 
                6.0 & $0.19 \pm 0.01$ & $0.60 \pm 0.01$ \\ 
                7.0 & $0.19 \pm 0.01$ & $0.66 \pm 0.01$ \\ 
                8.0 & $0.26 \pm 0.01$ & $0.75 \pm 0.01$ \\ 
                9.0 & $0.27 \pm 0.01$ & $0.74 \pm 0.01$ \\ 
                10.0 & $0.29 \pm 0.01$ & $0.76 \pm 0.02$ \\ 
                11.0 & $0.30 \pm 0.02$ & $0.80 \pm 0.02$ \\ 
                12.0 & $0.42 \pm 0.03$ & $0.97 \pm 0.05$ \\ 
                13.0 & $0.41 \pm 0.01$ & $1.15 \pm 0.03$ \\ 
                14.0 & $0.37 \pm 0.01$ & $1.20 \pm 0.02$ \\ 
                15.0 & $0.39 \pm 0.03$ & $1.45 \pm 0.07$ \\ 
                16.0 & $0.48 \pm 0.04$ & $1.93 \pm 0.17$ \\ 
                17.0 & $0.54 \pm 0.05$ & $2.39 \pm 0.23$ \\ 
                18.0 & $0.64 \pm 0.07$ & $2.63 \pm 0.34$ \\
                19.0 & $0.76 \pm 0.06$ & $3.35 \pm 0.50$ \\
                20.0 & $0.77 \pm 0.08$ & $2.77 \pm 0.92$ \\[1ex]  
                \hline
            \end{tabular}
        }
    	\caption{Dependence of the scale height of the Galactic disc as a function of the Galactocentric distance for the whole data sample and for Galactic azimuths $\phi \in \left[330^{\circ}, 30^{\circ}\right]$}\label{tab-comp2}
    	\label{tabulka}
\end{table}

\begin{figure}  
        \includegraphics[width=9cm]{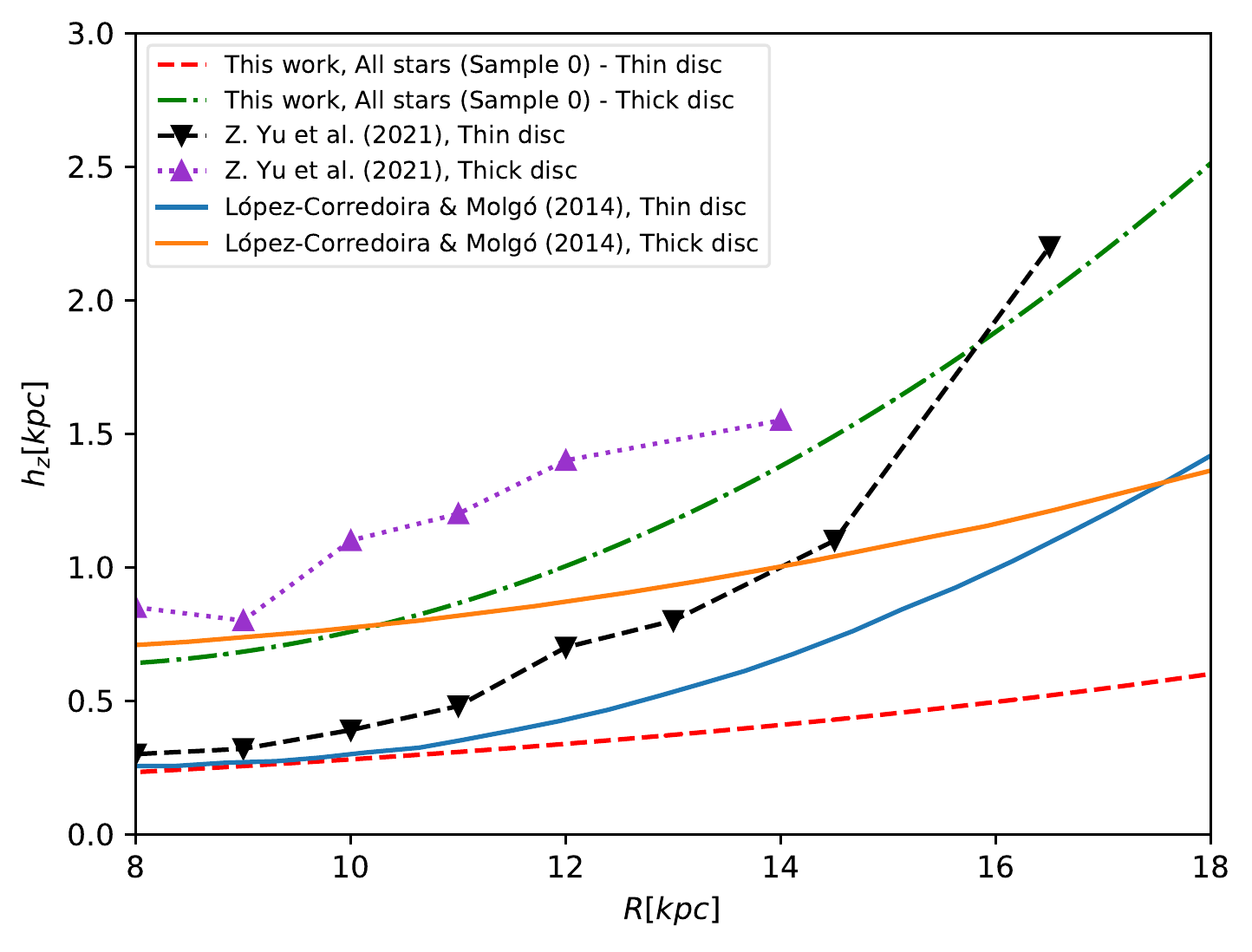}
        \caption{Comparison of the flare for the whole population (Sample 0) with other works. Our work is represented by the polynomial fits to the scale height data points (for more details and data points with error bars, see Tab.\ref{tab-comp2} and Fig. \ref{hz}).}\label{hzcomp1}
\end{figure}

The Galactic flare is significant in Sample 0, which represents the whole population. The scale  height of the thin disc rises from $0.26\pm0.01$ kpc in the solar neighbourhood ($R=8$ kpc) to $0.77\pm0.08$ kpc in the remote regions of the Galactic disc ($R=20$ kpc). The flaring of the thick disc is even more significant, where the scale height increases from $0.75\pm0.01$ kpc in $R=8$ kpc to reach $3.35\pm0.50$ kpc in $R=19$ kpc. The strong flare in the thick disc is present despite large error bars of $h_z$ for $R>16$ kpc. Comparison of the results with other works is presented in Fig. \ref{hzcomp1}. The flaring of the thin and the thick discs is presented in the results of all authors, but our data exhibits stronger flaring of the thick disc in the remote regions for $R>14$ kpc compared to \cite{yu} and \cite{martin_flare}.

\begin{figure}  
        \includegraphics[width=9cm]{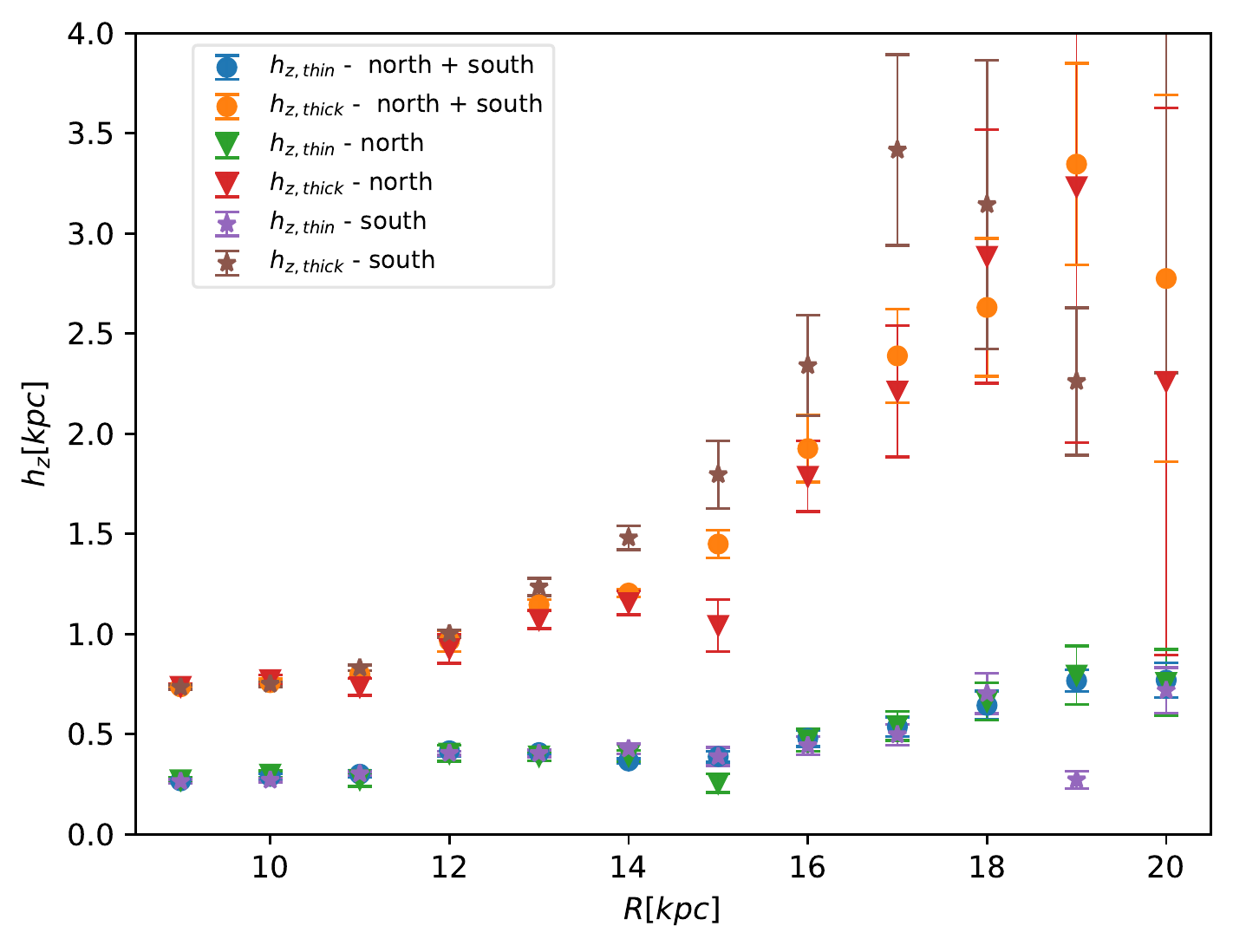}
        \caption{Dependence of the scale height of the thick and the thin discs on the Galactocentric distance. The Galactic azimuth is $\phi \in \left[330^{\circ}, 30^{\circ}\right]$. The northern, the southern, and the northern+southern flares are compared.}\label{figNS}
\end{figure}

\subsection{The northern and southern flare} \label{sec-NS}
We also focused on the differences between the northern and the southern flare. In Fig. \ref{figNS} we plot the comparison of the northern, southern and northern+southern flares. There is no significant difference in the dependence of the scale height with the Galactocentric distance for $R<15$ kpc. However, for larger distances, the flaring of the thick disc is asymmetric. The value of the southern scale height is approxomately $2$ kpc higher than the $h_z$ of the northern flare in $R=17$ kpc, although the error bars due to the lack of robust datasets in this region have to be taken into account. The difference decreases for $R>17$ kpc and the scale height error bars of the northern and southern flares overlap. 

\subsection{The azimuthal dependence of the scale height} \label{sec-azim}
\begin{figure}  
        \includegraphics[width=9cm]{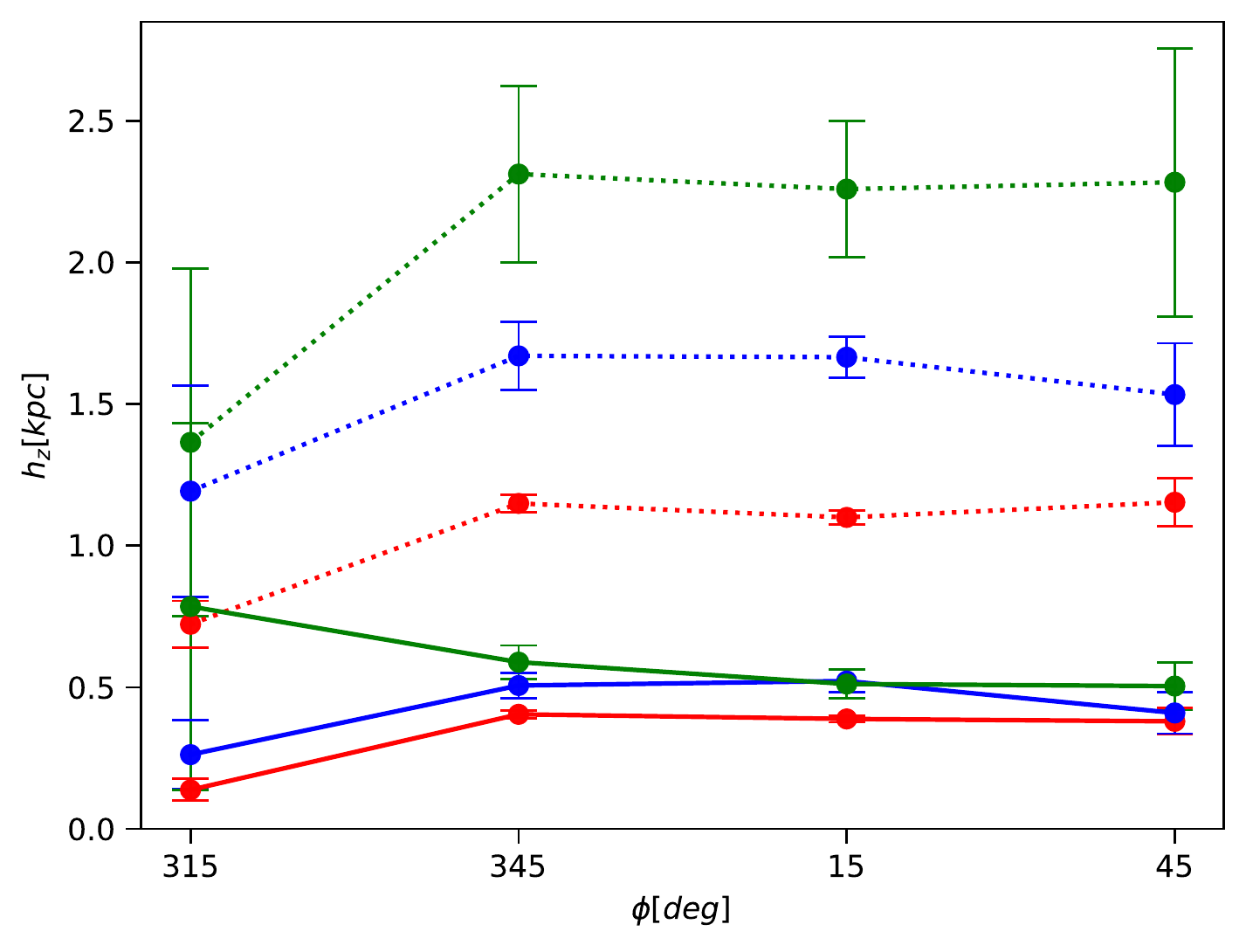}
        \caption{Dependence of the scale height on the Galactic azimuth $\phi$ for various Galactocentric distances: $R=13$ kpc (red lines); $R=15$ kpc (blue lines); $R=17$ kpc (green lines). Dotted lines represent the scale height of the thick disc and solid lines represent the scale height of the thin disc. Azimuth is binned with size $\Delta\phi=\ang{30}$.}\label{fig-azim}
\end{figure}

In order to study the azimuthal dependence of the scale height, we divided the Galactic disc into sectors. We used the following azimuth intervals  $\phi \in \left[300^{\circ}, 330^{\circ}\right]$, $\phi \in \left[330^{\circ}, 0^{\circ}\right]$, $\phi \in \left[0^{\circ}, 30^{\circ}\right]$, and $\phi \in \left[30^{\circ}, 60^{\circ}\right]$. The results for various Galactocentric distances $R \in [13,15,17]$ kpc are plotted in Fig. \ref{fig-azim}.

The dependence of the scale height is not significant in our data for $R<17$ kpc. The $h_z$ value of the thick disc in $R=17$ kpc for the azimuth $\Phi \in [300^\circ,330^\circ]$ is significantly lower than for the rest of azimuth intervals. However, considering the error bars of $h_z$, the azimuthal asymmetry almost vanishes.

\subsection{Supergiants} \label{sec-SG}
\begin{figure}  
        \includegraphics[width=9cm]{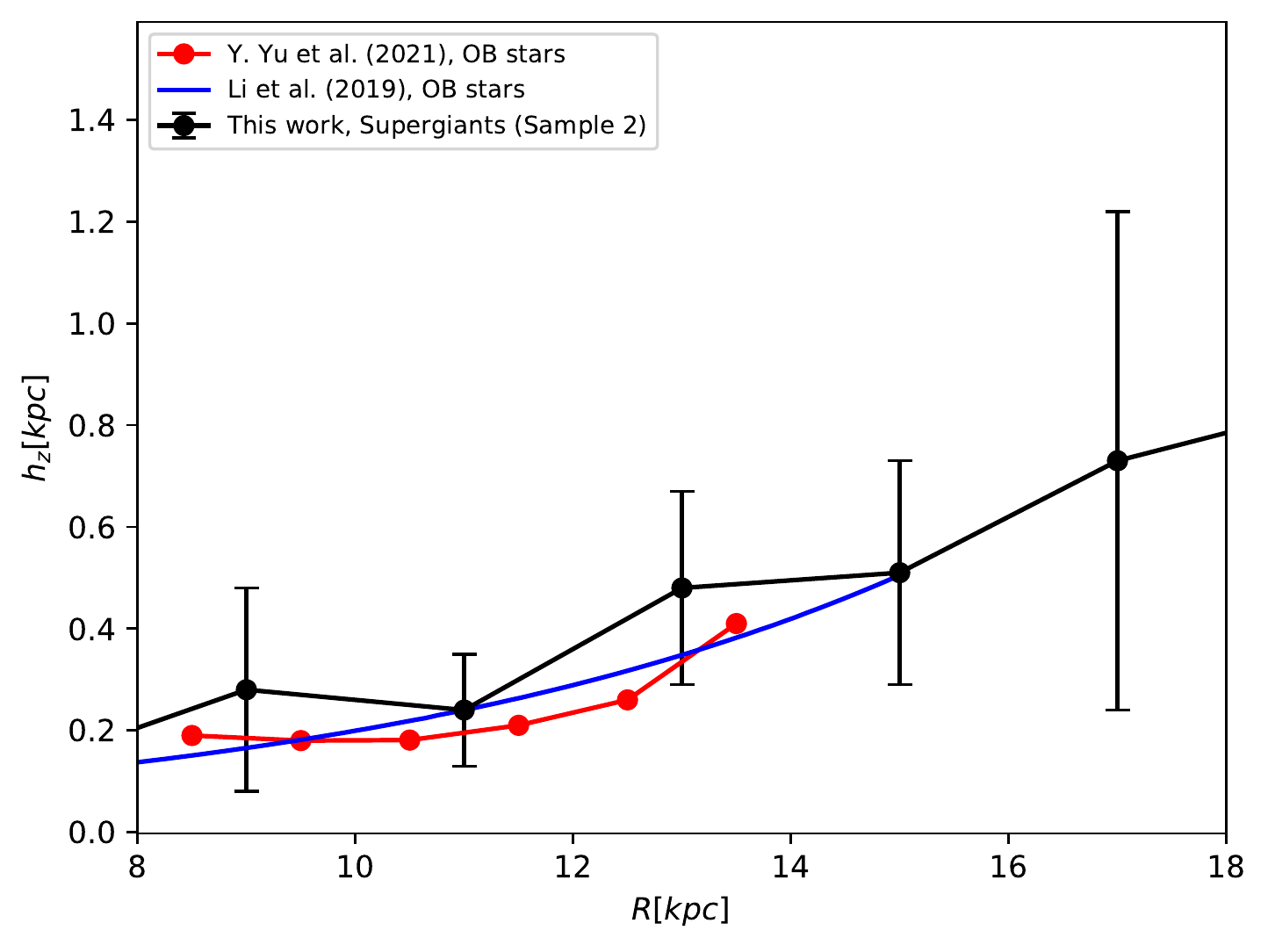}
        \caption{Comparison of the thin disc scale heights of the supergiants (Sample 2) with other works.}\label{hzcomp2}
\end{figure}

We also investigated the flaring of supergiants. We only used Sample 2, defined in Section \ref{sample2}, as Sample 1 was incomplete and therefore could not be used to model the flare. Since we only put constraints on the absolute magnitude of the stars ($-5<M<-10$), contamination is expected in the dataset, although as mentioned is Section \ref{warp}, it is not significant. We followed the flare analysis procedure described in Sections \ref{sec-meth} - \ref{sec-vertical}. We considered the thin disc and we did not put any constrains on the azimuth. Due to the significantly smaller size of the dataset, we used larger bins: $\Delta R = 2$ kpc and $\Delta z = 0.3$ kpc. We calculated the scale length $h_r = 1.99$ $\pm$ $0.13$ kpc of the disc. The flaring of Sample 2 stars is presented in Table \ref{tab-SG}. One can see the subtle increase in $h_z$, from values $h_z\approx 0.2$ kpc in the solar neighbourhood ($R\approx 8$ kpc) to $h_z \approx 0.8$ kpc in the remote regions of the Milky Way ($R\approx 18$ kpc). A significant increase in the scale height appears for $R\geq 13$ kpc, where $h_z \in [0.48, 0.84]$ kpc. Although the error bars are huge in this interval of the Galactocentric distances (from $\pm0.2$ kpc to $\pm0.7$ kpc), the higher value of the scale height for $R\in[13, 15]$ kpc ($hz \approx 0.5$ kpc) might be a real feature in the studied data sample. The comparison with other works using OB stars is plotted in Fig. \ref{hzcomp2}.

\begin{table}
        \centering
        \begin{tabular}{|c c|} 
        	\hline
        	R [kpc] & Scale-height [kpc] \\
        	\hline\hline
        	$7$ & $0.13 \pm 0.06$ \\ 
        	$9$ & $0.28 \pm 0.20$ \\
        	$11$ & $0.24 \pm 0.11$ \\
        	$13$ & $0.48 \pm 0.19$ \\
        	$15$ & $0.51 \pm 0.22$ \\
        	$17$ & $0.73 \pm 0.49$ \\
        	$19$ & $0.84 \pm 0.68$ \\[1ex] 
        	\hline
        \end{tabular}
        \caption{Scale height of the Galactic disc for the supergiants (Sample 2). The data is binned with size $\Delta z = 0.3$ kpc and $\Delta R = 2$ kpc.}
        \label{tab-SG}
\end{table}

\section{Conclusions}
We used Gaia EDR3 to study the outer Galactic disc using supergiants and compared them with the whole population. We concentrated on the Galactic warp and flare.
The warp of the whole population is similar to the results from previous works \citep[e.g.][]{zofi}, reaching a maximum amplitude of $z_w=0.360$ kpc and a minimum amplitude of $z_w=-0.375$ kpc at a distance $R=[19.5,20]$ kpc, revealing a small asymmetry between the northern and the southern warp. The warp of the supergiants, which are notably younger than the whole population, reaches a much larger maximum amplitude of $z_w=0.658$ kpc and a minimum of $z_w=-0.717$ kpc at a distance $R=[19.5,20]$ kpc, with the north-south asymmetry maintained. The difference between the warps of the two populations is significant, confirming a significant relationship between the age of the studied population and the warp amplitude. This result suggests that the warp is induced by a non-gravitational mechanism, such as accretion of intergalactic matter onto the disc or an intergalactic magnetic field.

We find a significant flare of the whole population, especially in the thick disc. The scale height increases from $h_{z,thick}=0.75 \pm 0.01$ kpc and $h_{z,thin} = 0.26 \pm 0.01$ kpc for $R = 8$ kpc, to $h_{z,thick} = 2.63\pm 0.34$ kpc and $h_{z,thin} = 0.64\pm0.07$ kpc at the Galactocentric distance $R\approx 18$ kpc), if the azimuth $\Phi \in [330^\circ,30^\circ]$ is considered. We also investigated the dependence of the scale height on the azimuth, which is not present for $R<17$ kpc. For $R=17$ kpc and $\Phi \in [300^\circ,330^\circ],$ the changes of the $h_z$ are visible. However, considering the error bars of the $h_z$, the azimuthal asymmetry almost vanishes. On the other hand, we find a small north-south asymmetry, especially in the thick disc. The asymmetry appears for $R>15$ kpc and the value of $h_z$ of the southern flare is approximately $1$ kpc higher than for the northern flare. However, the error bars of the $h_z$ for the northern and the southern flares overlap for $R>17$ kpc.
A subtle flare is present in the supergiants population, in comparison with the whole population of the disc. We find a significant increase in the $h_z$ for $R\geq 13$, but considering error bars, this rise of $h_z$ remains inconclusive. The $h_z$ error bars for $R\geq 13$ reach higher values due to a significant dispersion of densities in vertical profiles and asymmetry between northern and southern flares. It is therefore clear that the population of supergiants is warped more significantly than the whole population, while the flare for this population is less prominent.

Investigation into the warping and flaring mechanisms (e.g. various mechanisms of disc heating, mergers, magnetic field, etc.) is beyond the scope of this study. However, the forthcoming Gaia data releases promise significant improvement in positional and kinematic data, which will pave the way for a better understanding of the dynamics forming these structural features in the remote regions of the Milky Way.

\begin{acknowledgements}
We thank the anonymous referee for helpful comments, which improved this paper. ZC and MLC were supported by the grant PGC-2018-102249-B-100 of the Spanish Ministry of Economy and Competitiveness (MINECO). RN was supported by the VEGA - the Slovak Grant Agency for Science, grant No. 1/0761/21, by the Slovak Research and Development Agency under the contract No. APVV 18-0103 and by the Erasmus+ programe of the European Union under grant No. 2020-1-CZ01-KA203-078200. This work made use of the IAC Supercomputing facility HTCondor (http://research.cs.wisc.edu/htcondor/), partly financed by the Ministry of Economy and Competitiveness with FEDER funds, code IACA13-3E-2493. This work has made use of data from the European Space Agency (ESA) mission {\it Gaia} (\url{https://www.cosmos.esa.int/gaia}), processed by the {\it Gaia} Data Processing and Analysis Consortium (DPAC, \url{https://www.cosmos.esa.int/web/gaia/dpac/consortium}). Funding for the DPAC has been provided by national institutions, in particular the institutions participating in the {\it Gaia} Multilateral Agreement. 
\end{acknowledgements}

\DeclareRobustCommand{\disambiguate}[3]{#1}
\bibliographystyle{aa} 
\bibliography{edr3_supergiants_fin}

\end{document}